\documentclass[11pt,a4paper]{article}
\usepackage{authblk}
\usepackage{lmodern}
\usepackage{jheppub}
\usepackage{amsmath}
\usepackage{mathtools}
\usepackage{tikz}
\usepackage{empheq}
\usepackage{enumitem}
\usepackage{graphics}
\usepackage{wrapfig}
\usepackage{caption}
\usepackage{natbib}
\usepackage{xcolor}
\usepackage{framed}
\usepackage{array}
\usepackage{dsfont}
\definecolor{shadecolor}{gray}{0.925}

\usepackage[cbgreek]{textgreek}

\def\sideremark#1{\ifvmode\leavevmode\fi\vadjust{\vbox to0pt{\vss
 \hbox to 0pt{\hskip\hsize\hskip1em
 \vbox{\hsize3cm\tiny\raggedright\pretolerance10000
 \noindent #1\hfill}\hss}\vbox to8pt{\vfil}\vss}}}%

\newcommand{\bi}{\begin{itemize}}
\newcommand{\ei}{\end{itemize}}
\newcommand{\bea}{\begin{align}}
\newcommand{\eea}{\end{align}}
\newcommand{\be}{\begin{equation}}
\newcommand{\ee}{\end{equation}}

\makeatletter
\renewcommand*\env@matrix[1][\arraystretch]{%
  \edef\arraystretch{#1}%
  \hskip -\arraycolsep
  \let\@ifnextchar\new@ifnextchar
  \array{*\c@MaxMatrixCols c}}
\makeatother
\author[\ensuremath{a},\ensuremath{b}]{Lorenzo IACOBACCI}
\author[\ensuremath{a},\ensuremath{b}]{Charlotte SLEIGHT}
\author[\ensuremath{a},\ensuremath{b},\ensuremath{c}]{\quad Massimo TARONNA}

\affiliation[\ensuremath{a}]{Dipartimento di Fisica ``Ettore Pancini'', Universit\`a degli Studi di Napoli Federico II, \\Monte S. Angelo, Via Cintia, 80126 Napoli, Italy}

\affiliation[\ensuremath{b}]{INFN, Sezione di Napoli, Monte S. Angelo, Via Cintia, 80126 Napoli, Italy}

\affiliation[\ensuremath{c}]{Scuola Superiore Meridionale, Universit\`a degli Studi di Napoli Federico II,\\ Largo San Marcellino 10, 80138 Napoli, Italy}

\emailAdd{lorenzo.iacobacci@unina.it,charlotte.sleight@na.infn.it, massimo.taronna@unina.it}

\title{\centering \huge Celestial Holography Revisited II:\\ Correlators and K\"all\'en-Lehmann}

\abstract{In this work we continue the investigation of the extrapolate dictionary for celestial holography recently proposed in \cite{Sleight:2023ojm}, at both the perturbative and non-perturbative level. Focusing on scalar field theories, we give a complete set of Feynman rules for extrapolate celestial correlation functions and their radial reduction in the hyperbolic slicing of Minkowski space. We prove to all orders in perturbation theory that celestial correlators can be re-written in terms of corresponding Witten diagrams in EAdS which, in the hyperbolic slicing of Minkowski space, follows from the fact that the same is true in dS space. We then initiate the study of non-perturbative properties of celestial correlators, deriving the radial reduction of the K\"all\'en-Lehmann spectral representation of the exact Minkowski two-point function. We discuss the analytic properties of the radially reduced spectral function, which is a meromorphic function of the spectral parameter, and highlight a connection with the Watson-Sommerfeld transform.}

\begin{document}

\begin{flushright}    
\texttt{}
\end{flushright}

\maketitle

\newpage

\section{Introduction}\label{sec::Intro}

The AdS/CFT correspondence \cite{Maldacena:1997re,Gubser:1998bc,Witten:1998qj} is the prime working example of the holographic principle, according to which quantum gravity in anti-de Sitter (AdS) space is identified with a Conformal Field Theory (CFT) on the boundary at infinity. Although AdS/CFT represents a major advance in our understanding of Quantum Gravity, it is imperative to extend its success to scenarios closer to that of our Universe. To this end, many efforts have focused on anti-de Sitter's more realistic maximally symmetric cousins, de Sitter (dS) space ($\Lambda>0$) and Minkowski space ($\Lambda=0$). A key difficulty in extending AdS/CFT to such space-times lies in the fact that the boundaries at infinity are time-like (in dS) or null (in Minkowski) and therefore lack a standard notion of boundary time, and this obscures how holographic observables might encode consistent unitary time evolution of bulk physics. This is to be contrasted with the situation in AdS space where the boundary, being at spatial infinity, shares the time direction with the bulk and consistent time evolution is then encoded by the standard quantum mechanical time evolution of the boundary system. 

\vskip 4pt
The focus of this work is on holographic observables on the celestial sphere of Minkowski space. A promising approach to celestial holography \cite{Raclariu:2021zjz,Pasterski:2021rjz,McLoughlin:2022ljp,Pasterski:2021raf} has been to draw wisdom from S-matrix theory. Celestial amplitudes \cite{Pasterski:2016qvg,Pasterski:2017kqt} define holographic observables on the co-dimension two celestial sphere as a conformal basis for Minkowski scattering amplitudes, with the view that new understanding might be imported from the S-matrices directly. This approach has been particularly successful for massless particles owing to the fact that the corresponding geodesics reach null infinity. For massive particles, where the geodesics instead do not reach null infinity, the connection with the S-matrix is less transparent. It is interesting to note that a similar difference between massive and massless particles arises in AdS space. While massless AdS geodesics do reach the boundary at infinity, massive geodesics don't. Non-the-less, the holographic correlators in AdS-CFT are agnostic to this difference since they are off-shell.

\vskip 4pt
Motivated by this observation, another approach to celestial observables has been to take inspiration from holographic correlation functions in AdS-CFT. Following the seminal work of de Boer and Solodukhin \cite{deBoer:2003vf}, such correlators were defined in \cite{Sleight:2023ojm} by considering a hyperbolic foliation of Minkowski space. Bulk Minkowski time-ordered correlation functions are then extrapolated to the celestial sphere by taking the Mellin transform with respect to the radial direction of the slicing and the boundary limit in the hyperbolic directions, which is analogous to the extrapolate definition of boundary correlators in AdS-CFT. 

\vskip 4pt
The hyperbolic slicing foliates Minkowski space with constant curvature co-dimension 1 hypersurfaces naturally associated to the light-cone structure: Outside the light cone the foliating surfaces are co-dimension 1 de Sitter (dS) spaces while inside they are co-dimension 1 Euclidean anti-de Sitter (EAdS) spaces. The proposal of de Boer and Solodukhin \cite{deBoer:2003vf} is that one might then define holography on the celestial sphere by applying holography to each hyperbolic slice. In the context dS holography, it has recently been shown that the Feynman rules in the Bunch-Davies vacuum can be recast as Feynman rules for Witten diagrams in EAdS under analytic continuation, with the result that dS boundary correlators can be written in terms of corresponding Witten diagrams in EAdS.\footnote{It is well known that dS and Euclidean AdS are related by analytic continuation, and this has been exploited in various ways \cite{Maldacena:2002vr,McFadden:2009fg,Sleight:2019hfp,Sleight:2020obc,Sleight:2021plv,DiPietro:2021sjt,Bzowski:2023nef} to connect boundary correlation functions in dS and EAdS.} The hyperbolic slicing of Minkowski space thus suggests that a similar relationship with Witten diagrams might hold for celestial correlation functions defined in \cite{Sleight:2023ojm}.\footnote{For celestial amplitudes as defined in \cite{Pasterski:2016qvg,Pasterski:2017kqt} instead, see e.g. \cite{Cheung:2016iub,Lam:2017ofc,Casali:2022fro,PipolodeGioia:2022exe,Iacobacci:2022yjo,Melton:2023bjw} on relations with Witten diagrams.} This would open up the possibility to import the wealth of understanding of holographic correlators in AdS to celestial holography.

\vskip 4pt 
That celestial correlators might be reformulated in terms of EAdS Witten diagrams was already implied in \cite{Sleight:2023ojm}, under some assumptions. In particular, it was observed that contact diagram contributions to correlators on the $d$-dimensional celestial sphere are proportional to the corresponding contact Witten diagrams in EAdS$_{d+1}$. Under the assumption of consistent factorisation and completeness of Principal Series representations of the Lorentz group $SO(1,d+1)$, Minkowski exchange processes then decompose into a continuum of exchange Witten diagrams in EAdS$_{d+1}$ where the exchanged particles carry Principal Series representations. In this work we \emph{prove} that celestial correlators indeed admit such a decomposition into EAdS Witten diagrams to all orders in perturbation theory. In particular, by deriving the radial reduction of the Minkowski Feynman rules onto the extended unit hyperboloid, time-ordered propagators in the dS region of the hyperbolic slicing decompose in terms of time-ordered propagators of the dS hyperboloid. One can then recycle the results of \cite{Sleight:2020obc,Sleight:2021plv} relating Feynman rules for boundary correlators in dS and EAdS.

\vskip 4pt 
Celestial correlators as defined in \cite{Sleight:2023ojm} therefore have a similar structure to their AdS counterparts in the Euclidean regime, at least perturbatively. A natural next step is to explore their non-perturbative structure, which we initiate in the second part of this work. An important tool in non-perturbative QFT is the K\"all\'en-Lehmann spectral representation of exact two-point functions. Using the results of the first part of this work, we derive the radial reduction of the K\"all\'en-Lehmann representation onto the extended unit hyperboloid which can then be used to import methods from non-perturbative QFT to Celestial Correlators. We present an initial study of how the properties of the K\"all\'en-Lehmann spectral representation translate under radial reduction and give some examples, leaving a more thorough discussion to an upcoming work \cite{toappear}.

\vskip 4pt
The main results of this work can be summarised as:
\begin{itemize}
    \item A complete set of Feynman rules for celestial correlation functions as defined in \cite{Sleight:2023ojm} and their radial reduction onto the extended unit hyperboloid in the hyperbolic slicing of Minkowski space.
    \item A reformulation of the Feynman rules in the hyperbolic slicing in terms of Feynman rules for Witten diagrams in EAdS via analytic continuation, recycling the analytic continuations \cite{Sleight:2020obc,Sleight:2021plv} of propagators in dS to those in EAdS. This culminates in a set of rules that allows to immediately write down any given perturbative contribution to celestial correlators in terms of corresponding Witten diagrams in EAdS.
    \item The radial reduction of the K\"all\'en-Lehmann spectral representation for exact Minkowski two-point functions on the extended unit hyperboloid. The analyticity properties of the radial spectral function inherited from the original K\"all\'en-Lehmann spectral function are discussed. We highlight a connection between the radially reduced K\"all\'en-Lehmann representation and the expansion in spherical harmonics in Euclidean space.
\end{itemize}

This paper is organised as follows: In section \ref{sec::HSMINK} we review the hyperbolic slicing of Minkowski space and the celestial sphere. In section \ref{sec::CCfs}, after reviewing the definition \cite{Sleight:2023ojm} of celestial correlation functions, in section \ref{subsec::revdStoEAdS} we review the Feynman rules for boundary correlators in EAdS and dS, and the relation \cite{Sleight:2020obc,Sleight:2021plv} between the two via analytic continuation. In section \ref{subsec::celestFeyn} we give a complete set of Feynman rules for celestial correlation functions in scalar field theories and their radial reduction onto the extended unit hyperboloid. In section \ref{subsec::CelesttoEAdS} we give the reformulation of the celestial Feynman rules in terms of those for Witten diagrams in EAdS. In section \ref{sec::pertcalcs} we use this reformulation to compute celestial correlators perturbatively, in particular: contact diagrams in section \ref{subsec::pertcalccontact} and exchange diagrams in section \ref{subsec::pertcalcexch}. In section \ref{subsec::pertcalcrules} we give a set of rules to immediately express any given pertubative contribution to celestial correlation functions in terms of corresponding Witten diagrams in EAdS. In section \ref{sec::RRKL} we consider the radial reduction of the K\"all\'en-Lehmann spectral representation for exact Minkowski two-point functions. In section \ref{subsec::revKL} we review the standard K\"all\'en-Lehmann representation and in section \ref{subsec::RR} derive its radial reduction and discuss its properties. In section \ref{subsec::KLEX} we consider some perturbative examples. Various technical details are relegated to the appendices.

\newpage

\section{Hyperbolic slicing of Minkowski space}
\label{sec::HSMINK}

\begin{figure}[htb]
    \centering
    \includegraphics[width=0.6\textwidth]{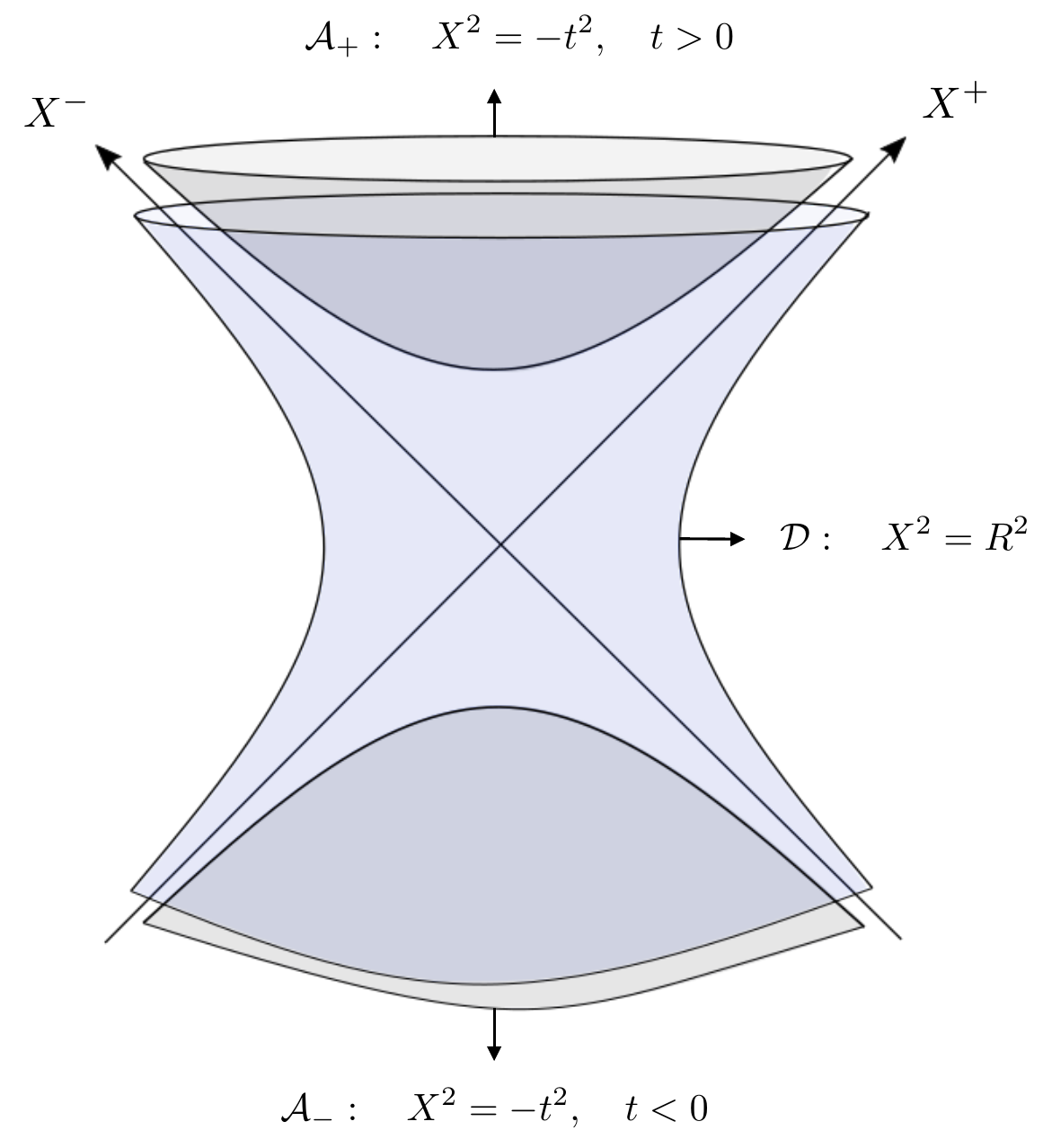}
    \caption{Hyperbolic foliation of Minkowski space with light-cone coordinates $X^\pm$.}
    \label{fig::HMink}
\end{figure}

Consider $\left(d+2\right)$-dimensional Minkowski space $\mathbb{M}^{d+2}$ with coordinates $X^M$, ${M=0,\ldots, d+1}$ and metric
\begin{equation}
    ds^2=-\left(dX^0\right)^2+\left(dX^1\right)^2+\ldots+\left(dX^{d+1}\right)^2. 
\end{equation} 
The light-cone $X^2=0$ naturally splits $\mathbb{M}^{d+2}$ into three regions:
\begin{subequations}
\begin{align}
    \mathcal{D}:&\quad X^2>0\,,\\
    \mathcal{A}_+:&\quad X^2<0\,,\qquad X^0>0,\\
    \mathcal{A}_-:&\quad X^2<0\,,\qquad X^0<0.
\end{align}
\end{subequations}
Analogous to a spherical foliation of $\mathbb{R}^{d+2}$, each of these regions can be foliated with surfaces of constant curvature which reflects the $SO\left(1,d+1\right)$ Lorentz symmetry. See figure \ref{fig::HMink}. In $\mathcal{A}_\pm$ these are $\left(d+1\right)$-dimensional Euclidean anti-de Sitter spaces (EAdS$_{d+1}$) with constant radius $t$:
\begin{equation}
    X^2=-t^2, \qquad t\in \mathbb{R}. \label{EAdSt}
\end{equation}
A useful set of coordinates for the hyperbolic foliation of ${\cal A}_\pm$, which is particularly well-suited for holography, is given by Poincar\'e coordinates
\begin{subequations}\label{EAdScoords}
\begin{align}
    \mathcal{A}_+:&\qquad X^M=+\frac{t}{z}\left(\frac{1+z^2+\vec{x}^2}{2},\frac{1-z^2-\vec{x}^2}{2},\vec{x}\right),& t&>0,\quad z>0,\\
    \mathcal{A}_-:&\qquad X^M=+\frac{t}{z}\left(\frac{1+z^2+\vec{x}^2}{2},\frac{1-z^2-\vec{x}^2}{2},\vec{x}\right),& t&<0,\quad z>0.
\end{align}
\end{subequations}
where $\vec{x}\in\mathbb{R}^d$. For region ${\cal D}$ instead, the foliating surfaces are $\left(d+1\right)$-dimensional de Sitter space-times with radius $R$:
\begin{equation}
    X^2=R^2. \label{dSr}
\end{equation}
Each dS space is covered by two Poincar\'e patches:
\begin{subequations}\label{dScoords}
\begin{align}
    \mathcal{D}_+:&\qquad X^M=\frac{R}{(-\eta)}\left(\frac{1-\eta^2+\vec{x}^2}{2},\frac{1+\eta^2-\vec{x}^2}{2},\vec{x}\right),& R&>0,\quad \eta<0\\
    \mathcal{D}_-:&\qquad X^M=-\frac{R}{\eta}\left(\frac{1-\eta^2+\vec{x}^2}{2},\frac{1+\eta^2-\vec{x}^2}{2},\vec{x}\right),& R&>0,\quad \eta>0 
\end{align}
\end{subequations}
which correspond to the expanding and contracting patches of the dS hypersurface \eqref{dSr} respectively:
\begin{subequations}
\begin{align}
    \mathcal{D}_+:&\quad X^2>0\,,\qquad X^+>0,\\
    \mathcal{D}_-:&\quad X^2>0\,,\qquad X^+<0,
\end{align}
\end{subequations}
where $X^+$ is the light cone coordinate $X^+=X^{0}+X^{1}$. 

\vskip 4pt
\paragraph{Conformal Boundary.} The conformal boundaries of each hyperbolic slice are identified with the projective cone of light rays (see figure \ref{fig::HMink})
\begin{equation}\label{PrLC}
    Q^2=0, \qquad Q \equiv \lambda Q, \qquad \lambda \in \mathbb{R}_+,
\end{equation}
which corresponds to $d$-dimensional spheres $S^-_d$ ($Q^0<0$) and $S^+_d$ ($Q^0>0$) in the infinite past and infinite future, respectively. This can be made manifest by introducing projective coordinates 
\begin{align}
    \xi_1 = Q^1/Q^0, \quad \xi_2 = Q^2/Q^0, \quad \ldots \quad, \quad \xi_{d+1} = Q^{d+1}/Q^0,
\end{align}
in terms of which the equation of the sphere is
\begin{equation}
    \xi^2_1+\ldots+\xi^2_{d+1}-1=0,
\end{equation}
which is the \emph{celestial sphere}. In the hyperbolic slicing of Minkowski space it is therefore natural to formulate flat space holography with boundary given by the celestial sphere, which was proposed by de Boer and Solodukhin in \cite{deBoer:2003vf}. In such a formulation, holography in Minkowski space can be thought of as being induced by applying holography to each hyperbolic slice.

\vskip 4pt
The regions ${\cal A}_+$ and ${\cal A}_-$ foliated by $\left(d+1\right)$-dimensional Euclidean anti-de Sitter spaces have conformal boundaries $S^+_d$ and $S^-_d$, respectively, at spatial infinity. In Poincar\'e coordinates \eqref{EAdScoords} these are reached by sending $z \to 0$ with parameterisation
\begin{subequations}
\begin{align}
    Q_+&\sim \frac{t}{z}\, \left(\frac{1+x^2}2,\frac{1-x^2}2,\vec{x}\right),&  &\text{with}&  t>0,\\
    Q_-&\sim \frac{t}{z}\,\left(\frac{1+x^2}2,\frac{1-x^2}2,\vec{x}\right),&  &\text{with}& t<0.
\end{align}
\end{subequations}
The region ${\cal D}$ foliated by $\left(d+1\right)$-dimensional de Sitter space-times has conformal boundary $\left(S^+_d\right)$ at future infinity  and $\left(S^-_d\right)$ past infinity. In Poincar\'e coordinates \eqref{dScoords} these are obtained in the limit $\eta \to 0$ with parameterisation
\begin{subequations}
\begin{align}
    Q_+&\sim \frac{R}{(-\eta)}\left(\frac{1+x^2}2,\frac{1-x^2}2,\vec{x}\right),&  &\text{with}&  \eta<0, \\Q_- &\sim -\frac{R}{\eta}\left(\frac{1+x^2}2,\frac{1-x^2}2,\vec{x}\right), &  &\text{with}&  \eta>0.
\end{align}
\end{subequations}

\section{Celestial Correlation functions}
\label{sec::CCfs}

In the presence of a translation symmetry, one often finds it is natural to work in Fourier space where the translation generator is diagonalised. In the hyperbolic slicing of Minkowski space however, translation symmetry is no longer manifest and it is instead natural to decompose into representations of $SO\left(1,d+1\right)$ i.e. the isometry group of each hyperbolic slice. These are labelled by the Eigenvalue $\Delta$ of the dilatation generator (a scaling dimension) and $SO\left(d\right)$ spin. The dilatation generator is diagonalised by the Mellin transform in the radial direction,
\begin{equation}
    \phi_{\Delta}({\hat X}) = \int^{\infty}_{0}\frac{dt}{t} t^{\Delta}\phi(t {\hat X}),
\end{equation}
where ${\hat X}$ parameterises the extended unit hyperboloid by writing $X = t {\hat X}$. This has inverse
\begin{equation}\label{phirr}
    \phi\left(X\right) = \int^{c+i\infty}_{c-i\infty}\frac{d\Delta}{2\pi i}\,t^{-\Delta}\,\phi_{\Delta}({\hat X}).
\end{equation}

\vskip 4pt
Based on this observation, in \cite{Sleight:2023ojm} holographic correlation functions on the Celestial sphere were defined as the boundary limit of Mellin transformed time-ordered bulk Minkowski correlation functions. Considering the correlation functions of $n$ scalar fields $\phi_i$, $i=1,\ldots,n$, in Minkowski space, the corresponding Celestial correlator is:
\begin{align}\label{ccdefn}
    \left\langle\mathcal{O}_{\Delta_1}(Q_1)\ldots \mathcal{O}_{\Delta_n}(Q_n)\right\rangle =\prod_i \lim_{{\hat X}_i\to Q_i}\,\int^\infty_0 \frac{dt_i}{t_i}\,t_i^{\Delta_i}\left\langle\phi_1(t_1\hat{X}_1)\ldots \phi_n(t_n\hat{X}_n)\right\rangle.
\end{align}
This definition naturally extends the extrapolate definition of holographic correlators in AdS and dS.

\vskip 4pt
In section \ref{subsec::celestFeyn} we give a complete set of Feynman rules for the perturbative computation of holographic celestial correlators as defined by \eqref{ccdefn}, focusing for simplicity on theories of scalar fields. In section \ref{subsec::CelesttoEAdS} we will show how such Celestial Feynman rules can be reformulated in terms of Feynman rules for boundary correlators in EAdS$_{d+1}$ of unit radius via appropriate analytic continuations in the hyperbolic slicing. To this end, we begin in the following section with a review of the perturbative computation of boundary correlation functions in EAdS and dS, and relation between the two. In the process we extend the latter relation beyond the expanding patch of dS.

\subsection{Review of (EA)dS Feynman rules}
\label{subsec::revdStoEAdS}

In this section we review the Feynman rules for scalar theories in Euclidean AdS and dS. It has been shown \cite{Sleight:2019mgd,Sleight:2019hfp,Sleight:2020obc,Sleight:2021plv} that the Feynman rules for boundary (Schwinger-Keldysh) correlators in the expanding patch of dS can be recast via analytic continuation as Feynman rules for boundary correlators in Euclidean AdS. In this section we will review these results and extend them to propagators in both the expanding and contracting patch, which will be needed to apply them to celestial correlators since both patches are present in the hyperbolic slicing of Minkowski space.

\vskip 4pt
Consider $\left(d+1\right)$-dimensional Euclidean anti-de Sitter space of unit radius, where the two sheets can be parameterised in Poincar\'e coordinates via
\begin{align}
    {\hat X}_{{\cal A}_\pm} = \pm \frac{ 1}{z}\left(\frac{1+z^2+\vec{x}^2}{2},\frac{1-z^2-\vec{x}^2}{2},\vec{x}\right), \qquad z >0.
\end{align}
Two point functions are functions of the EAdS chordal distance 
\begin{align}
    \sigma_{\text{AdS}}({\hat X},{\hat Y}) = \frac{1}{2}\left(1+{\hat X}\cdot {\hat Y}\right),
\end{align}
where the bulk-to-bulk propagators for a scalar field of mass $m_{\text{AdS}}$ are the Green's functions
\begin{equation}\label{bubuads}
    G^{\text{AdS}}_{\Delta}\left(\sigma_{\text{AdS}}\right) = C^{\text{AdS}}_{\Delta} \left(-4\sigma_{\text{AdS}}\right)^{-\Delta} {}_2F_1\left(\begin{matrix}\Delta,\Delta-\frac{d}{2}+\frac{1}{2}\\2\Delta-d+1\end{matrix},\frac{1}{\sigma_{\text{AdS}}}\right),
\end{equation}
with normalisation
\begin{equation}\label{ads2pt}
C^{\text{AdS}}_{\Delta}=\frac{\Gamma\left(\Delta\right)}{2\pi^{\frac{d}{2}}\Gamma\left(\Delta-\frac{d}{2}+1\right)}, 
\end{equation}
and $\Delta$ is a solution to the quadratic equation:
\begin{equation}
    m_{\text{AdS}}^2 = \Delta\left(d-\Delta\right).
\end{equation}
It is often useful to parameterise these solutions by $\Delta_\pm=\frac{d}{2}\pm i\mu$ where $\Im\left[\mu\right]<0$. The difference of the two solutions for the bulk-to-bulk propagator is a regular homogeneous solution of the Klein-Gordon equation,
\begin{align}\label{adsharm}
\Omega_{\mu}\left(\sigma_{\text{AdS}}\right) &= \frac{i\mu}{2\pi}\left[G^{\text{AdS}}_{\Delta_+}\left(\sigma_{\text{AdS}}\right)-G^{\text{AdS}}_{\Delta_-}\left(\sigma_{\text{AdS}}\right)\right],\\
    &=\frac{1}{(4\pi)^{\frac{d+1}2}\Gamma(\frac{d+1}2)}\frac{\Gamma(\tfrac{d}{2}+i\mu)\Gamma(\tfrac{d}{2}-i\mu)}{\Gamma(i\mu)\Gamma(-i\mu)} {}_2{F}_1\left(\begin{matrix}\tfrac{d}{2}+i\mu,\tfrac{d}{2}-i\mu\\\frac{d+1}{2}\end{matrix};\sigma_{\text{AdS}}\right).
\end{align}
Taking $\mu \in \mathbb{R}$ these provide a complete orthogonal basis for normalisable two-point functions in EAdS \cite{Cornalba:2007fs,Penedones:2010ue}, which is reviewed in appendix \ref{app::spectralrep}.

\vskip 4pt
The bulk-to-boundary propagator is obtained from the bulk-to-bulk propagator \eqref{bubuads} upon sending one of the bulk points to the boundary 
\begin{align}\label{buboads}
    K^{\text{AdS}}_{\Delta}\left(s_{\text{AdS}}\right)&=   \lim_{\lambda \to \infty} \lambda^{\Delta} G^{\text{AdS}}_{\Delta}\left({\hat X},{\hat Y}=\lambda Q + \ldots \right),\\
    &= \frac{C^{\text{AdS}}_{\Delta}}{\left(-2 s_{\text{AdS}} \right)^{\Delta}},
\end{align}
where the $\ldots$ serve to enforce the constraint ${\hat Y}^2=-1$ and $s_{\text{AdS}}({\hat X},Q) = {\hat X}\cdot Q$. In Poincar\'e coordinates this is
\begin{align}\label{AdSbubopoin}
K^{\text{AdS}}_{\Delta}\left(z,\vec{x};\vec{y}\right)=C^{\text{AdS}}_{\Delta} \left(\frac{z}{z^2+\left(\vec{x}-\vec{y}\right)^2}\right)^{\Delta},
\end{align}
where in ${\cal A}_\pm$ we parameterise the boundary points as
\begin{equation}\label{Qbdads}
    Q_\pm = \pm \left(\frac{1+y^2}2,\frac{1-y^2}2,\vec{y}\right).
\end{equation}

\vskip 4pt
Consider now $\left(d+1\right)$-dimensional de Sitter space of unit radius, where the expanding and contracting patches can be parameterised in Poincar\'e coordinates via 
\begin{align}
    {\hat X}_{{\cal D}_+} &= -\frac{1}{\eta}\left(\frac{1-\eta^2+\vec{x}^2}{2},\frac{1+\eta^2-\vec{x}^2}{2},\vec{x}\right), \qquad \eta <0,\\
    {\hat X}_{{\cal D}_-} &= -\frac{1}{\eta}\left(\frac{1-\eta^2+\vec{x}^2}{2},\frac{1+\eta^2-\vec{x}^2}{2},\vec{x}\right), \qquad \eta >0.
\end{align}
Two-point functions are now functions of the dS chordal distance
\begin{align}
    \sigma_{\text{dS}}\left({\hat X},{\hat Y}\right) &= \frac{1}{2}\left(1+{\hat X}\cdot {\hat Y}\right).
    \\
&=1+\frac{\left(\eta_1-\eta_2\right)^2-\left(\vec{x}_1-\vec{x}_2\right)^2}{4\eta_1 \eta_2}.
\end{align}
Propagators for a scalar field of mass $m_{\text{dS}}$ in the Bunch-Davies (Euclidean) vacuum of dS \cite{Gibbons:1977mu,Bunch:1978yq}\footnote{In dS there is in fact a one-parameter family as possible vacua that are invariant under the dS isometry group \cite{Allen:1985ux}. The Bunch-Davies vacuum can be uniquely defined as the one in which the Green's functions satisfy the Hadamard condition - that they behave as in Minkowski space on the light cone.} are derived from the function
\begin{equation}\label{homods}
   G_{\Delta}\left(\sigma\right) = \Gamma(i\mu)\Gamma(-i\mu) \Omega_{\mu}\left(\sigma\right),
\end{equation}
where now
\begin{equation}
    m^2_{\text{dS}} = \Delta \left(\Delta-d\right).
\end{equation}
Unlike in EAdS \eqref{adsharm}, where the function \eqref{homods} is regular, in dS it has a short distance singularity at $\sigma_{\text{dS}}=1$ and a branch cut for $\sigma_{\text{dS}}\in\left(1,\infty\right)$ where the two-points become time-like separated. Different prescriptions for approaching the branch cut give rise to the different dS two-point functions. The (anti)-Feynman propagators $G_{T}$ are given by
\begin{subequations}\label{(A)TdS}
  \begin{align}\label{TdS}
    G^{\text{dS}}_{\Delta, T}\left(\sigma_{\text{dS}}\right) &= G_{\Delta}\left(\sigma_{\text{dS}}-i \epsilon\right),\\
    G^{\text{dS}}_{\Delta, {\bar T}}\left(\sigma_{\text{dS}}\right) &= G_{\Delta}\left(\sigma_{\text{dS}}+i \epsilon\right).
\end{align}  
\end{subequations}
The corresponding bulk-to-boundary propagators can be derived similarly as in EAdS as a boundary limit of bulk-to-bulk propagators as in dS (see \cite{Sleight:2019mgd,Sleight:2019hfp}), giving 
\begin{subequations}\label{dSbubo(A)T}
 \begin{align}\label{dSbuboT}
K^{\text{dS}}_{\Delta,\,T}\left(s_{\text{dS}}\right)&= \frac{C^{\text{dS}}_{\Delta}}{\left(-2 s_{\text{dS}} + i \epsilon \right)^{\Delta}},\\
K^{\text{dS}}_{\Delta,\,{\bar T}}\left(s_{\text{dS}}\right)&= \frac{C^{\text{dS}}_{\Delta}}{\left(-2 s_{\text{dS}} - i \epsilon \right)^{\Delta}},
\end{align}   
\end{subequations}
where $s_{\text{dS}}({\hat X},Q) = {\hat X}\cdot Q$ and two-point coefficient
\begin{equation}
   C^{\text{dS}}_{\Delta}= \frac{1}{4\pi^{\frac{d+2}2}}\,\Gamma(\Delta)\Gamma(\tfrac{d}2-\Delta).
\end{equation}
In Poincar\'e coordinates, these read (for both expanding and contracting patches)
\begin{subequations}
     \begin{align}
K^{\text{dS}}_{\Delta,\,T}\left(\eta,\vec{x};\vec{y}\right)&= C^{\text{dS}}_{\Delta}\left(\frac{-\eta}{\mp\eta^2\pm\left(\vec{x}-\vec{y}\right)^2+i\epsilon}\right)^{\Delta},\\
K^{\text{dS}}_{\Delta,\,{\bar T}}\left(\eta,\vec{x};\vec{y}\right)&= C^{\text{dS}}_{\Delta}\left(\frac{-\eta}{\mp\eta^2\pm\left(\vec{x}-\vec{y}\right)^2-i\epsilon}\right)^{\Delta},
 \end{align}   
\end{subequations}
where the $\pm$ refer to the location of the boundary point $Q_\pm$ i.e. future $Q_+$ or past $Q_-$ boundary.

\vskip 4pt
The dS Feynman rules reviewed can be reformulated in EAdS via analytic continuation. In particular, it is well known that dS and EAdS are formally related by analytic continuation, which at the level of the Poincar\'e patches is implemented via $\eta \to \pm i z$. dS two-point functions are therefore a linear combination of analytically continued bulk-to-bulk propagators \eqref{bubuads} for a field of the same mass in EAdS, which is dictated by the $\epsilon$ prescription \cite{Sleight:2019mgd,Sleight:2019hfp}: 
\begin{align}\label{todsads}
time\text{-}ordering\,&: \quad \eta \to +i z,\\
anti\text{-}time\text{-}ordering\,&: \quad \eta \to -i z,
\end{align}
which in terms of embedding coordinates can be expressed as  
\begin{align}\label{todsadsT}
time\text{-}ordering\,&: \quad {\hat X}_{{\cal D}_\pm} \, \to \, \pm i {\hat X}_{{\cal A}_\pm},\\
anti\text{-}time\text{-}ordering\,&: \quad {\hat X}_{{\cal D}_\pm} \, \to \, \mp i {\hat X}_{{\cal A}_\pm},
\end{align}
where points in ${\cal D}_\pm$ continue to points in ${\cal A}_\pm$. Under these the chordal distance becomes
\begin{align}\label{contdpdp}
\sigma_{\text{dS}}({\hat X}_{{\cal D}_\pm},{\hat Y}_{{\cal D}_\pm}) &\to 1-\sigma_{\text{AdS}}({\hat X}_{{\cal A}_\pm},{\hat Y}_{{\cal A}_\pm}),\\
\sigma_{\text{dS}}({\hat X}_{{\cal D}_\pm},{\hat Y}_{{\cal D}_\mp}) &\to \sigma_{\text{AdS}}({\hat X}_{{\cal A}_\pm},{\hat Y}_{{\cal A}_\mp}).
\end{align}
In the case that one point lies in ${\cal D}_+$ and the other in ${\cal D}_-$, from the identity \eqref{adsharm} we immediately have that 
\begin{align}
    G^{\text{dS}}_{\Delta, T}({\hat X}_{{\cal D}_\pm},{\hat Y}_{{\cal D}_\mp}) \,&\to\, c^{\text{dS-AdS}}_{\Delta_+}G^{\text{AdS}}_{\Delta_+}(\sigma_{\text{AdS}}({\hat X}_{{\cal A}_\pm},{\hat Y}_{{\cal A}_\mp})-i\epsilon)\,+\,\left(\Delta_+\,\to\,\Delta_-\right),\\
    G^{\text{dS}}_{\Delta, {\bar T}}({\hat X}_{{\cal D}_\pm},{\hat Y}_{{\cal D}_\mp}) \,&\to\, c^{\text{dS-AdS}}_{\Delta_+}G^{\text{AdS}}_{\Delta_+}(\sigma_{\text{AdS}}({\hat X}_{{\cal A}_\pm},{\hat Y}_{{\cal A}_\mp})+i\epsilon)\,+\,\left(\Delta_+\,\to\,\Delta_-\right),
\end{align}
where the coefficient 
   \begin{equation}\label{dS-AdScoeff}
   c^{\text{dS-AdS}}_{\Delta} = \frac{C^{\text{dS}}_{\Delta}}{C^{\text{AdS}}_{\Delta}}=\frac{1}{2}\csc\left(\tfrac{\pi}{2}\left(d-2\Delta\right)\right),
   \end{equation}
accounts for the change in two-point coefficient from AdS to dS. Instead, when both bulk points in dS lie in the same patch, we have the identities \cite{Sleight:2020obc,Sleight:2021plv}: 
\begin{align}\label{analdSbubuT}
    G^{\text{dS}}_{\Delta, T}({\hat X}_{{\cal D}_\pm},{\hat Y}_{{\cal D}_\pm}) &\,\to\, c^{\text{dS-AdS}}_{\Delta_+}e^{-\Delta_+\pi i}G^{\text{AdS}}_{\Delta_+}({\hat X}_{{\cal A}_\pm},{\hat Y}_{{\cal A}_\pm})+\,\left(\Delta_+ \to \Delta_-\right),\\ \label{analdSbubuAT}
    G^{\text{dS}}_{\Delta, {\bar T}}({\hat X}_{{\cal D}_\pm},{\hat Y}_{{\cal D}_\pm}) &\,\to\, c^{\text{dS-AdS}}_{\Delta_+}e^{+\Delta_+\pi i}G^{\text{AdS}}_{\Delta_+}({\hat X}_{{\cal A}_\pm},{\hat Y}_{{\cal A}_\pm})+\,\left(\Delta_+ \to \Delta_-\right).
    \end{align}

\vskip 4pt
Analogous relationships can be obtained for bulk-to-boundary propagators either by taking the boundary limit or comparing the bulk-to-boundary propagators in dS \eqref{dSbubo(A)T} and EAdS \eqref{buboads}:
\begin{subequations}
\begin{align}\label{analdSbuboT}
K^{\text{dS}}_{\Delta,\,T}({\hat X}_{{\cal D}_\pm},Q)&\to  c^{\text{dS-AdS}}_{\Delta}\,i^{\mp \Delta}\,K^{\text{AdS}}_{\Delta}(s_{\text{AdS}}({\hat X}_{{\cal A}_\pm},Q)-i\epsilon),\\
K^{\text{dS}}_{\Delta,\,{\bar T}}({\hat X}_{{\cal D}_\pm},Q)&\to  c^{\text{dS-AdS}}_{\Delta}\,i^{\pm \Delta}\,K^{\text{AdS}}_{\Delta}(s_{\text{AdS}}({\hat X}_{{\cal A}_\pm},Q)+i\epsilon).
\end{align}   
\end{subequations}
Bulk integrals over dS space are recast as integrals over a sheet of EAdS under the analytic continuations \eqref{todsads}: 
\begin{align}\label{analtintdS}
    time\text{-}ordered\,&: \quad \int_{{\cal D}_\pm} d^{d+1}{\hat X} \, \to \, e^{\pm \frac{d \pi i}{2}}\int_{{\cal A}_\pm} d^{d+1}{\hat X},\\
anti\text{-}time\text{-}ordered\,&: \quad \int_{{\cal D}_\pm} d^{d+1}{\hat X} \, \to \, e^{\mp \frac{d \pi i}{2}}\int_{{\cal A}_\pm} d^{d+1}{\hat X}.
\end{align}

\vskip 4pt
With the above we have shown how to recast Feynman rules for scalar fields in dS as Feynman rules for scalar fields of the same mass in EAdS,\footnote{Strictly speaking one should also consider the analytic continuation of interaction vertices, where derivative interactions introduce additional phases under \eqref{todsads}. In this work however we restrict for simplicity to non-derivative interactions.} extending the results of \cite{Sleight:2019hfp,Sleight:2020obc,Sleight:2021plv} to include the contracting patch of dS.

\subsection{Feynman rules for Celestial Correlators}
\label{subsec::celestFeyn}

In this section we give the Feynman rules for the perturbative computation of Celestial correlation functions as defined via \eqref{ccdefn}. Minkowski two-point functions can be decomposed in terms of two-point functions on the extended unit hyperboloid where, depending on the region of Minkowski space, each bulk point is either a point in dS or EAdS space. This is detailed in the following. The Feynman rules can then be recast in terms of propagators in EAdS using the dS to EAdS map reviewed in the previous section, which is carried out in section \ref{subsec::CelesttoEAdS}.

\vskip 4pt
The \emph{celestial bulk-to-bulk propagator} is simply the free theory Minkowski space time-ordered two-point function (i.e. the Feynman propagator), which decomposes in Mellin space as follows
\begin{equation}\label{bubucelest}
G^{\left(m\right)}_{T}\left(X_1,X_2\right)=\int^{c+i\infty}_{c-i\infty}\frac{d\Delta_1}{2\pi i}\frac{d\Delta_2}{2\pi i}\, G^{(m)}_{\Delta_1 \Delta_2}({\hat X}_1,{\hat X}_2)\,t^{-\Delta_1}_1\,t^{-\Delta_2}_2,
\end{equation}
where (appendix \ref{app::doublemellinbubu}) 
\begin{align}\label{bubucelestM}
    G^{(m)}_{\Delta_1 \Delta_2}({\hat X}_1,{\hat X}_2) &= \int^{\infty}_{0} \frac{dt_1}{t_1} \frac{dt_2}{t_2}\,t^{\Delta_1}_1\,t^{\Delta_2}_2\,G^{\left(m\right)}_{T}(t_{1}{\hat X}_1,t_{2}{\hat X}_2),\\
    &=\frac{1}{2}\frac{1}{(4\pi)^{\frac{d+1}2}}\frac{m^{d-\Delta_1-\Delta_2}}{\left(\sqrt{\hat{X}^2_1+i\epsilon}\right)^{\Delta_1}\left(\sqrt{\hat{X}^2_2+i\epsilon}\right)^{\Delta_2}}\\&\times\frac{\Gamma(\tfrac{\Delta_1+\Delta_2-d}{2})
 \Gamma (\Delta_1) \Gamma (\Delta_2)}{\Gamma\left(\frac{\Delta_1+\Delta_2+1}{2}\right)}  \, _2F_1\left(\begin{matrix}\Delta_1,\Delta_2\\\frac{\Delta_1+\Delta_2+1}{2}\end{matrix};\sigma_\epsilon\right)\,\nonumber\\
 \sigma_\epsilon&=\frac{1}{2}\left(1-\frac{-\hat{X}\cdot\hat{Y}+i\epsilon}{\sqrt{\hat{X}^2+i\epsilon}\sqrt{\hat{Y}^2+i\epsilon}}\right).
\end{align}
This admits the spectral decomposition 
\begin{align}\label{spectralMT2pt}
G_{\Delta_1,\Delta_2}(\hat{X},\hat{Y})=\int_{-\infty}^{+\infty}\frac{d\nu}{2\pi}\, \rho^{(m)}_{\Delta_1,\Delta_2}(\nu)G_{\frac{d}{2}+i\nu}(\sigma_\epsilon)\,,
\end{align}
with
\begin{multline}
   t_1^{-\Delta_1-\frac{d}{2}}t_2^{-\Delta_2-\frac{d}{2}}\,\rho^{(m)}_{\Delta_1+\frac{d}{2},\Delta_2+\frac{d}{2}}(\nu)=\left(\frac{m}{2}\right)^{-\Delta_1-\Delta_2}\frac{1}{8\left(\sqrt{{X}^2+i\epsilon}\right)^{\Delta_1+\frac{d}{2}}\left(\sqrt{{Y}^2+i\epsilon}\right)^{\Delta_2+\frac{d}{2}}}\\\times\frac{\Gamma \left(\frac{\Delta_1-i \nu}{2}\right) \Gamma \left(\frac{\Delta_1+i \nu}{2}\right) \Gamma \left(\frac{\Delta_2-i \nu}{2}\right) \Gamma \left(\frac{\Delta_2+i \nu}{2}\right)}{\Gamma(+i\nu)\Gamma(-i\nu)}\,,\nonumber
\end{multline}    
which follows from completeness and orthogonality of the EAdS harmonic functions \eqref{adsharm}.

\vskip 4pt
The spectral decomposition \eqref{spectralMT2pt} of the Mellin transformed Feynman propagator gives a radial reduction of the position space Minkowski Feynman propagator onto the extended unit hyperboloid. In particular, performing the inverse Mellin transform \eqref{bubucelest} it gives rise to 
\begin{align}\label{spectralrepcelesybubu}
G^{(m)}_T(X,Y)=\int_{-\infty}^{+\infty}\frac{d\nu}{2\pi}\, \rho^{(m)}_{\nu}\left(\sqrt{X^2+i\epsilon},\sqrt{Y^2+i\epsilon}\right)G_{\frac{d}{2}+i\nu}(\sigma_\epsilon)\,,
\end{align}
with 
\begin{subequations}\label{SDFeyn}
\begin{align}
    \rho^{(m)}_{\nu}\left(\sqrt{X^2+i\epsilon},\sqrt{Y^2+i\epsilon}\right)&=\int^{c+i\infty}_{c-i\infty}\frac{d\Delta_1}{2\pi i}\frac{d\Delta_2}{2\pi i}\,t_1^{-\Delta_1-\frac{d}{2}}t_2^{-\Delta_2-\frac{d}{2}}\,\rho^{(m)}_{\Delta_1+\frac{d}{2},\Delta_2+\frac{d}{2}}(\nu),\,\\
    &=\frac{1}{2}\,\mathcal{K}_\nu^{(m)}(\sqrt{X^2+i\epsilon})\mathcal{K}_{-\nu}^{(m)}(\sqrt{Y^2+i\epsilon}),
\end{align}
\end{subequations}
and
\begin{align}\label{Kkernel}
        \mathcal{K}^{(m)}_{\nu}(R)&=\left(\frac{m}{2}\right)^{-i\nu}\frac{R^{-\frac{d}{2}}}{\Gamma\left(-i\nu\right)}\int^{c+i\infty}_{c-i\infty}\frac{d\Delta}{2\pi i}\Gamma\left(\Delta+\tfrac{i\nu}{2}\right)\Gamma\left(\Delta-\tfrac{i\nu}{2}\right)\left(\tfrac{mR}{2}\right)^{-2s},\\
    &=\left(\frac{m}{2}\right)^{-i\nu}\frac{2 R^{-\frac{d}{2}}}{\Gamma(-i\nu)}\,K_{i\nu}(m R)\,,
\end{align}
where $K_{\nu}(z)$ is the modified Bessel function of the second kind. In regions ${\cal D}_\pm$, which are foliated by dS$_{d+1}$ space-times, the decomposition \eqref{spectralrepcelesybubu} is a superposition of Feynman propagators in dS$_{d+1}$. In particular, we have
\begin{align}\label{TdS2}
    \sigma_{\epsilon}({\hat X}_{{\cal D}_\pm},{\hat Y}_{{\cal D}_{\hat \pm}})=  \sigma_{\text{dS}}({\hat X}_{{\cal D}_\pm},{\hat Y}_{{\cal D}_{\hat \pm}})-i \epsilon,
\end{align}
so that, via \eqref{TdS},
\begin{equation}\label{GTsuptdS}
   G^{(m)}_T(X_{{\cal D}_\pm},Y_{{\cal D}_{\hat \pm}})= \frac{1}{2} \int_{-\infty}^{+\infty}\frac{d\nu}{2\pi}\,\mathcal{K}_\nu^{(m)}(R_1)\mathcal{K}_{-\nu}^{(m)}(R_2)G^{\text{dS}}_{\frac{d}{2}+i\nu,\, T}({\hat X}_{{\cal D}_\pm},{\hat Y}_{{\cal D}_{\hat \pm}}).
\end{equation}
In regions ${\cal A}_\pm$ foliated by Euclidean AdS$_{d+1}$ spaces, we have 
\begin{align}
    \sigma_{\epsilon}({\hat X}_{{\cal A}_\pm},{\hat Y}_{{\cal A}_{\hat \pm}})=  1-\sigma_{\text{AdS}}({\hat X}_{{\cal A}_\pm},{\hat Y}_{{\cal A}_{\hat \pm}})+i \epsilon.
\end{align}
When both points belong to the same sheet of EAdS we can then use the identity \eqref{analdSbubuAT} to obtain:
\begin{multline}
G^{(m)}_T(X_{{\cal A}_\pm},Y_{{\cal A}_\pm})=\int_{-\infty}^{+\infty}\frac{d\nu}{2\pi}\,e^{\left(\frac{d}{2}+i\nu\right)\pi i}c^{\text{dS-AdS}}_{\frac{d}{2}+i\nu}\,\mathcal{K}_\nu^{(m)}(\pm e^{\frac{\pi i}{2}}t_1)\mathcal{K}_{-\nu}^{(m)}(\pm e^{\frac{\pi i}{2}}t_2)\,\\ \times G^{\text{AdS}}_{\frac{d}{2}+i\nu}({\hat X}_{{\cal A}_\pm},{\hat Y}_{{\cal A}_\pm}).
\end{multline}
In summary, the radial reduction \eqref{spectralrepcelesybubu} of the Minkowski Feynman propagator is a superposition of Greens functions on the extended unit hyperboloid carrying Principal Series representations of the isometry group $SO\left(1,d+1\right)$. In the de Sitter regions ${\cal D}_\pm$ these are time-ordered two-point functions on dS$_{d+1}$, which is is inherited from the time-ordering of the original Minkowski Feynman propagator (as expected).

\vskip 4pt
The \emph{celestial bulk-to-boundary propagator} is the boundary limit of the Mellin transform of one of the bulk points \cite{Sleight:2023ojm}:
    \begin{align}\label{celestialbubo}
    G^{\left(m\right)}_{\Delta}\left(X,Q\right) &=\lim_{\hat{Y}\to Q} \int^\infty_0 \frac{dt}{t} t^\Delta\,G^{\left(m\right)}_T\left(X,t\hat{Y}\right),\\ \nonumber
    &=c_\Delta^{\text{dS-AdS}} \,\mathcal{K}^{(m)}_{i\left(\frac{d}{2}-\Delta\right)}\left(\sqrt{X^2+i\epsilon}\right)\,G^{\text{AdS}}_\Delta(X_\epsilon,Q).
\end{align}
The dependence on the hyperbolic directions is given by the corresponding (analytically continued) bulk-to-boundary propagator in anti-de Sitter space
\begin{align}\label{adsanalbubo}
    G^{\text{AdS}}_\Delta(X_\epsilon,Q)&=
   C^{\text{AdS}}_\Delta \frac{\left(\sqrt{X^2+i\epsilon}\right)^{\Delta}}{(-2 X\cdot Q+i\epsilon)^\Delta}\,.
   \end{align}
Notice that, as anticipated above, in the dS region ${\cal D}_\pm$ of the hyperbolic slicing this dependence is precisely that of the corresponding time-ordered bulk-to-boundary propagator in dS$_{d+1}$:
    \begin{align}\label{adsanalbuboDpm}
    G^{\left(m\right)}_{\Delta}\left(X_{{\cal D}_\pm},Q\right) &= \mathcal{K}^{(m)}_{i\left(\frac{d}{2}-\Delta\right)}\left(R\right)\,K^{\text{dS}}_{\Delta,\,T}(s_{\text{dS}}({\hat X}_{{\cal D}_\pm},Q)).
\end{align}
In the EAdS regions ${\cal A}_\pm$, in Poincar\'e coordinates \eqref{EAdScoords} we have
\begin{align} 
    \sqrt{X^2+i \epsilon}\,\Big|_{{\cal A}_+}&= e^{\frac{\pi i}{2}} t, \qquad \hspace*{0.2cm}  
    \sqrt{X^2+i \epsilon}\,\Big|_{{\cal A}_-} = e^{\frac{\pi i}{2}} |t|, 
\end{align}
so that in these regions: 
\begin{align}\label{cbuboApm}
    G^{\left(m\right)}_{\Delta}(X_{{\cal A}_\pm },Q)&=c_\Delta^{\text{dS-AdS}} \,\mathcal{K}^{(m)}_{i\left(\frac{d}{2}-\Delta\right)}( \pm e^{\frac{\pi i}{2}} t\,)\, i^ {+\Delta}K^{\text{AdS}}_{\Delta}\left(  s_{\text{AdS}}({\hat X}_{{\cal A}_\pm},Q)-i\epsilon\right),
\end{align}
in terms of the corresponding bulk-to-boundary propagator \eqref{buboads} in EAdS$_{d+1}$, where the $\epsilon$ prescription here is accounting for the fact that $Q$ can parameterise either boundary $S^\pm_d$.

\vskip 4pt
The free theory two-point function on the celestial sphere, is the boundary limit of the Mellin transformed Feynman propagator \eqref{bubucelestM} (see appendix \ref{app::doublemellinbubu}): 
\begin{align}\label{celestial2pt}
     G^{(m)}_{\Delta_1 \Delta_2}(Q_1,Q_2)&:= \lim_{{\hat X}_i \to Q_i}\,G^{(m)}_{\Delta_1 \Delta_2}({\hat X}_1,{\hat X}_2),\\
     &=\frac{C^{\text{flat}}_{\Delta_1}}{(-2Q_1\cdot Q_2+i\epsilon)^{\Delta_1}}(2\pi )\delta(i(\Delta_1-\Delta_2))\,, \nonumber
\end{align}
with normalisation 
\begin{equation}\label{celest2ptnorm}
   C^{\text{flat}}_{\Delta}= \left(\frac{m}2\right)^{d-2\Delta}\frac{1}{4\pi^{\frac{d+2}2}}\,\Gamma(\Delta)\Gamma(\Delta-\tfrac{d}2).
\end{equation}
This can be equivalently obtained as the boundary limit of the Mellin transformed of the celestial bulk-to-boundary propagator \eqref{celestialbubo}:
\begin{equation}
    G^{(m)}_{\Delta_1 \Delta_2}(Q_1,Q_2)=\lim_{{\hat Y}\to Q_2}\int^{\infty}_0\frac{dt}{t}t^{\Delta_2}G^{\left(m\right)}_{\Delta_1}\left(t{\hat Y},Q_1\right),
\end{equation}
as presented in \cite{Sleight:2023ojm}. 

\vskip 4pt
In the view of recasting celestial correlation functions in terms of EAdS Witten diagrams in later sections, it will be useful to introduce the ratio of celestial \eqref{celest2ptnorm} and AdS \eqref{ads2pt} boundary two-point function coefficients,
\begin{equation}\label{flatads2pt}
    c^{\text{flat-AdS}}_{\Delta}=\frac{C^{\text{flat}}_{\Delta}}{C^{\text{AdS}}_{\Delta}} = \left(\frac{m}2\right)^{d-2\Delta}\frac{\Gamma (\Delta -\tfrac{d}{2}) \Gamma (\Delta-\tfrac{d}{2}+1)}{2\pi},
\end{equation}
which is the celestial analogue of the coefficient \eqref{dS-AdScoeff} in dS.

\subsection{Reformulation in Euclidean AdS}
\label{subsec::CelesttoEAdS}

In this section we show that the Feynman rules for Celestial correlation functions presented in the previous section can be reformulated in Euclidean AdS$_{d+1}$ via appropriate analytic continuations for points in regions ${\cal D}_\pm$ of the hyperbolic slicing. In particular, in the previous section we observed that points on the dS hyperboloid in these regions are time-ordered and can therefore be mapped to EAdS$_{d+1}$ according to the analytic continuation \eqref{todsadsT} of time-ordered points in dS.

\vskip 4pt
For the celestial bulk-to-boundary propagator \eqref{adsanalbuboDpm} in regions ${\cal D}_\pm$, applying analytic continuation \eqref{analdSbuboT} we have 
\begin{align}
    G^{\left(m\right)}_{\Delta}(X_{{\cal D}_\pm},Q)&= \mathcal{K}^{(m)}_{i\left(\frac{d}{2}-\Delta\right)}\left(R\right)\,K^{\text{dS}}_{\Delta,\,T}(s_{\text{dS}}({\hat X}_{{\cal D}_\pm},Q))\\&\to c^{\text{dS-AdS}}_{\Delta}\,\mathcal{K}^{(m)}_{i\left(\frac{d}{2}-\Delta\right)}(R\,)i^{\mp \Delta}\,K^{\text{AdS}}_{\Delta}(s_{\text{AdS}}({\hat X}_{{\cal A}_\pm},Q)-i\epsilon),\label{cbuboDpm}
\end{align}
in terms of the corresponding bulk-to-boundary propagator \eqref{buboads} in EAdS$_{d+1}$.

\vskip 4pt
For Celestial bulk-to-bulk propagators \eqref{bubucelest} the starting point is the radial reduction \eqref{spectralrepcelesybubu}, which we repeat below for convenience:
\begin{equation}\label{celestialbubu}
G^{(m)}_T(X,Y)=\frac{1}{2}\int_{-\infty}^{+\infty}\frac{d\nu}{2\pi}\,\mathcal{K}_\nu^{(m)}(\sqrt{X^2+i\epsilon})\mathcal{K}_{-\nu}^{(m)}(\sqrt{Y^2+i\epsilon})G_{\frac{d}{2}+i\nu}(\sigma_\epsilon)\,,
\end{equation}
where
\begin{equation}
\sigma_\epsilon=\frac{1}{2}\left(1-\frac{-\hat{X}\cdot\hat{Y}+i\epsilon}{\sqrt{\hat{X}^2+i\epsilon}\sqrt{\hat{Y}^2+i\epsilon}}\right).
\end{equation}
As observed in the previous section, when both bulk points lie in ${\cal D}_\pm$, this is a superposition \eqref{GTsuptdS} of time-ordered bulk-to-bulk propagators \eqref{dSbuboT} in de Sitter space. We can therefore use the analytic continuations \eqref{analdSbubuT} to obtain
\begin{align}
   G^{(m)}_T(X_{{\cal D}_\pm},Y_{{\cal D}_{\hat \pm}})&= \frac{1}{2} \int_{-\infty}^{+\infty}\frac{d\nu}{2\pi}\,\mathcal{K}_\nu^{(m)}(R_1)\mathcal{K}_{-\nu}^{(m)}(R_2)G^{\text{dS}}_{\frac{d}{2}+i\nu,\, T}({\hat X}_{{\cal D}_\pm},{\hat Y}_{{\cal D}_{\hat \pm}})\\ \nonumber
   &\to \int_{-\infty}^{+\infty}\frac{d\nu}{2\pi}\,e^{-(\frac{d}{2}+i\nu)\pi i}c^{\text{dS-AdS}}_{\frac{d}{2}+i\nu}\mathcal{K}_\nu^{(m)}(R_1)\mathcal{K}_{-\nu}^{(m)}(R_2)G^{\text{AdS}}_{\frac{d}{2}+i\nu}({\hat X}_{{\cal A}_\pm},{\hat Y}_{{\cal A}_{\hat \pm}}),
\end{align}
as a superposition of bulk-to-bulk propagators in EAdS$_{d+1}$.

\vskip 4pt
One proceeds in a similar fashion when just one of the two bulk points is in either of the regions ${\cal D}_\pm$. For example, when one point is in ${\cal D}_+$ and the other in ${\cal A}_+$, we have
\begin{align}\label{sigepsDApmanal}
    \sigma_{\epsilon}({\hat X}_{{\cal D}_+},{\hat Y}_{{\cal A}_+}) &\to  \sigma_{\text{AdS}}({\hat X}_{{\cal A}_+},{\hat Y}_{{\cal A}_+}),
\end{align}
in which case we can use identity \eqref{adsharm} to obtain
\begin{align}
    G^{(m)}_T(X_{{\cal D}_+},Y_{{\cal A}_+})&\to \int_{-\infty}^{+\infty}\frac{d\nu}{2\pi}\,c^{\text{dS-AdS}}_{\frac{d}{2}+i\nu}\mathcal{K}_\nu^{(m)}(R_1)\mathcal{K}_{-\nu}^{(m)}(t_2e^ {\frac{\pi i}{2}})G^{\text{AdS}}_{\frac{d}{2}+i\nu}({\hat X}_{{\cal A}_+},{\hat Y}_{{\cal A}_+}).
\end{align}
For points in ${\cal D}_-$, an interesting observation, which will be made use of later on, is that under the analytic continuation \eqref{todsadsT} of ${\hat X}_{{\cal D}_-}$ we have 
\begin{equation}\label{DmAmsigma}
    \sigma_\epsilon\left({\hat X}_{{\cal D}_-},\cdot\right)\, \to \,\sigma_\epsilon\left({\hat X}_{{\cal A}_-},\cdot\right),
\end{equation}
so that the hyperbolic dependence on points in ${\cal D}_-$ is the same as that in ${\cal A}_-$ upon analytic continuation. More explicitly, we have:
\begin{multline}
    \sigma_\epsilon\left({\hat X}_{{\cal D}_-},{\hat Y}_{{\cal A}_\pm}\right)=\frac{1}{2}\left(1+i\left(-{\hat X}_{{\cal D}_-} \cdot {\hat Y}_{{\cal A}_\pm}+i\epsilon\right)\right)\\ \to \frac{1}{2}\left(1-{\hat X}_{{\cal A}_-} \cdot {\hat Y}_{{\cal A}_\pm}+i\epsilon\right)=\sigma_\epsilon\left({\hat X}_{{\cal A}_-},{\hat Y}_{{\cal A}_\pm}\right),
\end{multline}
\begin{multline}
    \sigma_\epsilon\left({\hat X}_{{\cal D}_-},{\hat Y}_{{\cal D}_\pm}\right)=\frac{1}{2}\left(1-\left(-{\hat X}_{{\cal D}_-} \cdot {\hat Y}_{{\cal D}_\pm}+i\epsilon\right)\right)\\ \to \frac{1}{2}\left(1+i\left(-{\hat X}_{{\cal A}_-} \cdot {\hat Y}_{{\cal D}_\pm}+i\epsilon\right)\right)=\sigma_\epsilon\left({\hat X}_{{\cal A}_-},{\hat Y}_{{\cal D}_\pm}\right).
\end{multline}

\vskip 4pt
To conclude the re-writing of the Feynman rules for celestial correlators in EAdS, the bulk integrals over Minkowski space decompose into integrals over regions ${\cal A}_\pm$ and ${\cal D}_\pm$ in the hyperbolic slicing:
\begin{equation}
    \int d^{d+2}X = \int_{{\cal A}_+} d^{d+2}X+\int_{{\cal A}_-} d^{d+2}X+\int_{{\cal D}_+} d^{d+2}X+\int_{{\cal D}_-} d^{d+2}X,
\end{equation}
where, for regions ${\cal A}_\pm$, we have
\begin{align}
    \int_{{\cal A}_+}d^{d+2}X &= \int^{\infty}_0 dt\,t^{d+1}\,\int_{{\cal A}_+} d^{d+1}{\hat X},\\
    \int_{{\cal A}_-}d^{d+2}X &= \int^{0}_{-\infty} dt\,\left(-t\right)^{d+1}\,\,\int_{{\cal A}_-} d^{d+1}{\hat X}.
\end{align}
The integrals over regions ${\cal D}_\pm$ can each be reformulated as integrals over ${\cal A}_\pm$ using the analytic continuations \eqref{analtintdS} for time-ordered points: 
\begin{align}
\int_{{\cal D}_+}d^{d+2}X \quad & \to \quad e^{+\frac{d\pi i}{2}}\int_{{\cal A}_+}d^{d+2}X,\\
        \int_{{\cal D}_-}d^{d+2}X \quad &\to \quad e^{-\frac{d\pi i}{2}}\int_{{\cal A}_-}d^{d+2}X. 
\end{align}

\vskip 4pt
With the above we have shown that celestial correlation functions \eqref{ccdefn} can be recast in terms of propagators on both sheets of EAdS at any order in perturbation theory. In fact, it turns out that an even stronger result holds: The contributions from regions ${\cal D}_-$ and ${\cal A}_-$ precisely cancel - so that celestial correlation functions are the sum of contributions from regions ${\cal D}_+$ and ${\cal A}_+$ \emph{only}.\footnote{This was observed in \cite{Sleight:2023ojm} for contact diagrams and in the present work we provide a proof for all orders in perturbation theory.} This is simplest to see following some examples in the next section, and has the implication that perturbation theory for celestial correlation functions can be reformulated as a set of Feynman rules for boundary correlation functions on a single sheet (${\cal A}_+$) of EAdS with unit radius.

\section{Perturbative Calculations}
\label{sec::pertcalcs}

In this section we apply the Feynman rules from section \ref{subsec::CelesttoEAdS} to re-write perturbative celestial correlation functions in terms of corresponding Witten diagrams in EAdS$_{d+1}$, combining the contributions from each region of Minkowski space in the hyperbolic slicing. In section \ref{subsec::pertcalccontact} we consider contact diagrams and in section \ref{subsec::pertcalcexch} four-point exchange diagrams. This culminates in section \ref{subsec::pertcalcrules} with a set of rules that allow to immediately determine the Witten diagram decomposition of any given perturbative process contributing to a celestial correlator.

\subsection{Contact diagrams}
\label{subsec::pertcalccontact}

Consider the contact interaction governed by the following vertex in $\mathbb{M}^{d+2}$
\begin{equation}
    {\cal V}_{12 \ldots n} = g \phi_1 \phi_2 \ldots \phi_n,
\end{equation}
where $\phi_i$ are real scalar fields with masses $m_i$, $i = 1, 2, \ldots, n$. The corresponding celestial contact diagram was computed in \cite{Sleight:2023ojm}, where it was found to be proportional to the corresponding contact diagram in Euclidean anti-de Sitter space, with proportionality factors accounting for the difference in two- and three-point coefficients. In the following we reproduce this result using the analytic continuations derived in section \ref{subsec::CelesttoEAdS}.

\vskip 4pt
The celestial contact diagram decomposes into contributions from each region ${\cal A}_\pm$ and ${\cal D}_\pm$ in the hyperbolic slicing of $\mathbb{M}^{d+2}$
\begin{align}\label{ccontact}
    -ig \int d^{d+2}X\, \prod^n_{i=1}\,G^{(m_i)}_{\Delta_i}\left(X,Q_{i}\right) = -ig\left(I_{{\cal A}_+}+I_{{\cal A}_-}+I_{{\cal D}_+}+I_{{\cal D}_-}\right),
\end{align}
where we defined
\begin{equation}
I_{\bullet}=\int_{\bullet} d^{d+2}X\, \prod^n_{i=1}\,G^{(m_i)}_{\Delta_i}\left(X,Q_i\right),
\end{equation}
with $\bullet = {\cal A}_+,\,{\cal A}_-,\,{\cal D}_+,\,{\cal D}_-$. 

\vskip 4pt
In regions ${\cal A}_\pm$, inserting the expressions \eqref{cbuboApm} for the celestial bulk-to-boundary propagators, we have
\begin{multline}
I_{{\cal A}_\pm} = \left(\prod^n_{i=1}i^{\Delta_i}c^{\text{dS-AdS}}_{\Delta_i}\right)R^{({\cal A})}_{\Delta_1 \ldots \Delta_n}\left(m_1,\ldots,m_n\right)\\ \times \int_{{\cal A}_\pm}d^{d+1}{\hat X}\prod^n_{i=1}\, K^{\text{AdS}}_{\Delta}\left(s_{\text{AdS}}({\hat X},Q_i)-i\epsilon\right).
\end{multline}
The contribution from the radial direction in these regions is encoded in the function 
 \begin{equation}\label{RA}
   R^{({\cal A})}_{\Delta_1 \ldots \Delta_n}\left(m_1,\ldots,m_n\right)=\int^{\infty}_{0}dR\,R^{d+1}\prod^n_{i=1}\,\mathcal{K}^{(m_i)}_{i\left(\frac{d}{2}-\Delta_i\right)}(e^{\frac{\pi i}{2}}R),
\end{equation}
which moreover encodes all dependence on the mass of the original fields in $\mathbb{M}^{d+2}$.

\vskip 4pt
In regions ${\cal D}_\pm$ instead, inserting the analytic continuations \eqref{cbuboDpm} for the celestial bulk-to-boundary propagators, the contributions in these regions can be reformulated in EAdS as 
\begin{multline}
I_{{\cal D}_\pm} = e^{\pm\frac{d\pi i}{2}}\left(\prod^n_{i=1}i^{\mp\Delta_i}c^{\text{dS-AdS}}_{\Delta_i}\right)R^{({\cal D})}_{\Delta_1 \ldots \Delta_n}\left(m_1,\ldots,m_n\right)\\ \times \int_{{\cal A}_\pm }d^{d+1}{\hat X}\prod^n_{i=1}\, K^{\text{AdS}}_{\Delta}\left(s_{\text{AdS}}({\hat X},Q_i)-i\epsilon\right),
\end{multline}
 where in this case the contribution from the radial direction is encoded by the function
 \begin{equation}\label{RD}
   R^{({\cal D})}_{\Delta_1 \ldots \Delta_n}\left(m_1,\ldots,m_n\right)=\int^{\infty}_{0}dR\,R^{d+1}\prod^n_{i=1}\,\mathcal{K}^{(m_i)}_{i\left(\frac{d}{2}-\Delta_i\right)}(R).
\end{equation}

Notice that $I_{{\cal D}_\pm}$ and $I_{{\cal A}_\pm}$ are proportional to the same integral over the hyperbolic directions ${\hat X}$. In fact, the same is true for their radial contributions, which differ by a phase:
\begin{equation}\label{RAtoRD}
  R^{({\cal A})}_{\Delta_1 \ldots \Delta_n}\left(m_1,\ldots,m_n\right)=e^{-\left(d+2\right)\frac{\pi i}{2}}\,R^{({\cal D})}_{\Delta_1 \ldots \Delta_n}\left(m_1,\ldots,m_n\right),
\end{equation}
which can be seen simply by rotating the integration contour in \eqref{RA}.

\vskip 4pt
Putting everything together, as anticipated, the contributions to the bulk integral from regions ${\cal A}_-$ and ${\cal D}_-$ cancel: 
\begin{equation}\label{AmDm0cont}
    I_{{\cal A}_-}+I_{{\cal D}_-}=0.
\end{equation}
We will see that this feature carries over to processes involving particle exchanges. The Celestial contact diagram \eqref{ccontact} is therefore the sum of contributions from regions ${\cal A}_+$ and ${\cal D}_+$, which can be combined to give
\begin{align}
    -ig \int d^{d+2}X\, \prod^n_{i=1}\,G^{(m_i)}_{\Delta_i}\left(X,Q_i\right) &= -ig\left(I_{{\cal A}_+}+I_{{\cal D}_+}\right),\\ \nonumber
    &=\left(\prod^n_{i=1}c^{\text{dS-AdS}}_{\Delta_i}\right)2\sin\left(\left(-d+\sum\limits^n_{i=1}\Delta_i\right)\frac{\pi}{2}\right) \\& \times
    R^{({\cal D})}_{\Delta_1 \ldots \Delta_n}\left(m_1,\ldots,m_n\right)\\& \times  \underbrace{g\int_{{\cal A}_+}d^{d+1}{\hat X}\prod^n_{i=1}\, K^{\text{AdS}}_{\Delta}\left(\,  s_{\text{AdS}}({\hat X},Q_i)-i\epsilon\right)}_{\text{Contact Witten diagram in EAdS$_{d+1}$}}. \nonumber
\end{align}
For later purposes it will be useful to write the Celestial contact diagram in the form:
\begin{multline}\label{adstoflatcontact}
    -ig \int d^{d+2}X\, \prod^n_{i=1}\,G^{(m_i)}_{\Delta_i}\left(X,Q_i\right) = g\,c^{\text{flat-AdS}}_{\Delta_1 \ldots \Delta_n}\left(m_1,\ldots,m_n\right)D_{\Delta_1 \ldots \Delta_n}\left(Q_1,\ldots,Q_n\right),
\end{multline}
in terms of the well-known EAdS D-function \cite{DHoker:1999kzh} 
\begin{equation}
    D_{\Delta_1 \ldots \Delta_n}\left(Q_1,\ldots,Q_n\right)=\int_{{\cal A}_+}d^{d+1}{\hat X}\prod^n_{i=1}\, K^{\text{AdS}}_{\Delta}\left(\,  s_{\text{AdS}}({\hat X},Q_i)-i\epsilon\right),
\end{equation}
which is appropriately extended along the complexified null cone according to the $i\epsilon$ prescription (appendix D of \cite{Sleight:2023ojm}). This is multiplied by
\begin{multline}\label{adstoflatnpt}
    c^{\text{flat-AdS}}_{\Delta_1 \ldots \Delta_n}\left(m_1,\ldots,m_n\right)=\left(\prod^n_{i=1}c^{\text{dS-AdS}}_{\Delta_i}\right) 2\sin\left(\left(-d+\sum\limits^n_{i=1}\Delta_i\right)\frac{\pi}{2}\right) \\ \times
    R^{({\cal D})}_{\Delta_1 \ldots \Delta_n}\left(m_1,\ldots,m_n\right).
\end{multline}
In summary, we see that celestial contact diagrams are proportional to corresponding contact Witten diagrams in EAdS. In the proportionality factor, the contribution from the radial direction is encoded by the function \eqref{RD} and the sinusoidal factor arises from combining the contributions from each of this regions ${\cal D}_\pm$ and ${\cal A}_\pm$, which have differing phases. Interestingly, as noted in \cite{Sleight:2023ojm}, the contribution from the radial direction is, up to normalisation, given by the \emph{same} Witten diagram but for the shadow fields in EAdS$_{-d+1}$ and in momentum space in the boundary directions, where the masses $m_i$ play the role of the modulus of the boundary momentum and the kernel \eqref{Kkernel} the role of the bulk-to-boundary propagator \cite{Freedman:1998tz}.\footnote{In particular, with these identifications, \eqref{RD} takes the form of the multiple Bessel-K integrals \cite{Bzowski:2013sza} in momentum space of CFT$_{-d}$.}

\vskip 4pt
In summary, in this section we recovered the result of \cite{Sleight:2023ojm}, which was obtained by directly integrating over $\mathbb{M}^{d+2}$/the regions ${\cal D}_\pm$ and ${\cal A}_\pm$, thus providing a check of the analytic continuations derived in section \ref{subsec::CelesttoEAdS}.

\subsection{Four-point exchange diagram}
\label{subsec::pertcalcexch}

In this section we consider the celestial correlation function corresponding to the four-point exchange diagram mediated by cubic vertices
\begin{equation}
    {\cal V}_{12 \phi} = g_{12}\, \phi_1 \phi_2 \phi, \qquad {\cal V}_{34 \phi} = g_{34}\, \phi_3 \phi_4 \phi,
\end{equation}
with external scalar fields $\phi_i$ of mass $m_i$ and the field $\phi$ has mass $m$.

\vskip 4pt
The corresponding celestial correlator reads
\begin{multline}\label{celestialexch}
 \hspace*{-0.75cm}   \left(-ig_{12}\right)\left(-ig_{34}\right) \int d^{d+2}X\,\int d^{d+2}Y\,G^{(m_1)}_{\Delta_1}\left(X,Q_1\right)G^{(m_2)}_{\Delta_2}\left(X,Q_2\right)G^{\left(m\right)}_{T}\left(X,Y\right)G^{(m_3)}_{\Delta_3}\left(Y,Q_3\right)G^{(m_4)}_{\Delta_4}\left(Y,Q_4\right)\\
    = \left(-ig_{12}\right)\left(-ig_{34}\right)\sum\limits_{\pm\,{\hat \pm}}\left(I_{{\cal A}_\pm\,{\cal A}_{\hat \pm}}+I_{{\cal A}_\pm\,{\cal D}_{\hat \pm}}+I_{{\cal D}_\pm\,{\cal A}_{\hat \pm}}+I_{{\cal D}_\pm\,{\cal D}_{\hat \pm}}\right),
\end{multline}
where the bulk integrals decompose into contributions from each region in the hyperbolic slicing of Minkowski space:
\begin{multline}
    I_{\bullet_1\,\bullet_2} = \int_{\bullet_1} d^{d+2}X\,\int_{\bullet_2} d^{d+2}Y\,G^{(m_1)}_{\Delta_1}\left(X,Q_1\right)G^{(m_2)}_{\Delta_2}\left(X,Q_2\right)\\ \times G^{\left(m\right)}_{T}\left(X,Y\right)G^{(m_3)}_{\Delta_3}\left(Y,Q_3\right)G^{(m_4)}_{\Delta_4}\left(Y,Q_4\right),
\end{multline}
where $\bullet_{1,2} = {\cal A}_+,\,{\cal A}_-,\,{\cal D}_+,\,{\cal D}_-$.

\vskip 4pt
As for the contact diagrams in the previous section, by applying the Feynman rules as formulated in section \ref{subsec::CelesttoEAdS}, each contribution can be expressed as a superposition of exchanges in AdS carrying Principal Series representations of the Lorentz group. Focusing first on the contributions from regions ${\cal A}_+$ and ${\cal D}_+$, we have
\begin{multline}\label{Iaa}
    I_{{\cal A}_+\,{\cal A}_+} = i^{\Delta_1+\Delta_2+\Delta_3+\Delta_4}\left(\,\prod^4_{i=1}c^{\text{dS-AdS}}_{\Delta_i}\right) \int^{\infty}_{-\infty} \frac{d\nu}{2\pi} \,e^{\left(\frac{d}{2}+i\nu\right)\pi i}c^{\text{dS-AdS}}_{\frac{d}{2}+i\nu}\, \\ \times R^{({\cal A})}_{\Delta_1 \Delta_2\,  \frac{d}{2}+i\nu}\left(m_{1,2},m\right)R^{({\cal A})}_{\frac{d}{2}-i\nu \, \Delta_3 \Delta_4}\left(m,m_{3,4}\right)\\ \times  {\cal A}^{\text{AdS}}_{\Delta_1,\Delta_2|\frac{d}{2}+i\nu|\Delta_3,\Delta_4}\left(Q_1,Q_2,Q_3,Q_4\right),
\end{multline}
\begin{multline}\label{Iad}
   I_{{\cal A}_+\,{\cal D}_+} = e^{+\frac{d\pi i}{2}}i^{\Delta_1+\Delta_2-\Delta_3-\Delta_4}\left(\,\prod^4_{i=1}c^{\text{dS-AdS}}_{\Delta_i}\right) \int^{\infty}_{-\infty} \frac{d\nu}{2\pi} \, c^{\text{dS-AdS}}_{\frac{d}{2}+i\nu}\,\\ \times R^{({\cal A})}_{\Delta_1 \Delta_2\,  \frac{d}{2}+i\nu}\left(m_{1,2},m\right)R^{({\cal D})}_{\frac{d}{2}-i\nu \, \Delta_3 \Delta_4}\left(m,m_{3,4}\right)\\ \times {\cal A}^{\text{AdS}}_{\Delta_1,\Delta_2|\frac{d}{2}+i\nu|\Delta_3,\Delta_4}\left(Q_1,Q_2,Q_3,Q_4\right),
\end{multline}
\begin{multline}\label{Ida}
  I_{{\cal D}_+\,{\cal A}_+} = e^{+\frac{d\pi i}{2}}i^{-\Delta_1-\Delta_2+\Delta_3+\Delta_4}\left(\,\prod^4_{i=1}c^{\text{dS-AdS}}_{\Delta_i}\right) \int^{\infty}_{-\infty} \frac{d\nu}{2\pi} \, c^{\text{dS-AdS}}_{\frac{d}{2}+i\nu} \\ \times R^{({\cal D})}_{\Delta_1 \Delta_2  \frac{d}{2}+i\nu}\left(m_{1,2},m\right)R^{({\cal A})}_{\frac{d}{2}-i\nu \Delta_3 \Delta_4}\left(m,m_{3,4}\right)\\ \times {\cal A}^{\text{AdS}}_{\Delta_1,\Delta_2|\frac{d}{2}+i\nu|\Delta_3,\Delta_4}\left(Q_1,Q_2,Q_3,Q_4\right),
\end{multline}
\begin{multline}\label{Idd}
   I_{{\cal D}_+\,{\cal D}_+} = e^{+d\pi i}i^{-\Delta_1-\Delta_2-\Delta_3-\Delta_4}\left(\,\prod^4_{i=1}c^{\text{dS-AdS}}_{\Delta_i}\right) \int^{\infty}_{-\infty} \frac{d\nu}{2\pi} \, e^{-\left(\frac{d}{2}+i\nu\right)\pi i}\, c^{\text{dS-AdS}}_{\frac{d}{2}+i\nu}\,\\ \times  R^{({\cal D})}_{\Delta_1 \Delta_2  \frac{d}{2}+i\nu}\left(m_{1,2},m\right)R^{({\cal D})}_{\frac{d}{2}-i\nu \Delta_3 \Delta_4}\left(m,m_{3,4}\right) \\ \times {\cal A}^{\text{AdS}}_{\Delta_1,\Delta_2|\frac{d}{2}+i\nu|\Delta_3,\Delta_4}\left(Q_1,Q_2,Q_3,Q_4\right),
\end{multline}
where  
\begin{multline}
    {\cal A}^{\text{AdS}}_{\Delta_1,\Delta_2|\frac{d}{2}+i\nu|\Delta_3,\Delta_4}\left(Q_1,Q_2,Q_3,Q_4\right)=\int_{{\cal A}_+} d^{d+1}{\hat X}\,\int_{{\cal A}_+} d^{d+1}{\hat Y}\,G^{\text{AdS}}_{\frac{d}{2}+i\nu}({\hat X},{\hat Y})\\ \times \prod^2_{i=1}K^{\text{AdS}}_{\Delta_i}\left(s_{\text{AdS}}({\hat X},Q_i)-i\epsilon\right)\prod^4_{i=3}K^{\text{AdS}}_{\Delta_i}\left(s_{\text{AdS}}({\hat Y},Q_i)-i\epsilon\right),
\end{multline}
is the exchange of a particle with scaling dimension $\Delta=\frac{d}{2}+i\nu$ in AdS. As for contact diagrams, the $\epsilon$-prescription accounts for the fact that boundary points $Q_i$ can be null separated and the radial dependence from each bulk point is encoded in the functions \eqref{RD} and \eqref{RA}.

\vskip 4pt
The contributions from the remaining regions cancel in a pair-wise fashion, where the contribution from a bulk point in region ${\cal D}_-$ cancels with the contribution from the same bulk point in region ${\cal A}_-$:
\begin{align}
    I_{\bullet {\cal D}_-}+I_{\bullet {\cal A}_-}&=0,\\
    I_{{\cal D}_- \bullet}+I_{{\cal A}_- \bullet}&=0.
\end{align}
For example, applying the Feynman rules we have
\begin{multline}
  I_{{\cal D}_-\,{\cal D}_-} = e^{-\frac{d\pi i}{2}}e^{-\frac{d\pi i}{2}}i^{\Delta_1+\Delta_2+\Delta_3+\Delta_4}\left(\prod^4_{i=1}c^{\text{dS-AdS}}_{\Delta_i}\right) \int^{\infty}_{-\infty} \frac{d\nu}{2\pi} \,  e^{-\left(\frac{d}{2}+i\nu\right)\pi i}\, c^{\text{dS-AdS}}_{\frac{d}{2}+i\nu} \\ \times R^{({\cal D})}_{\Delta_1 \Delta_2  \frac{d}{2}+i\nu}\left(m_{1,2},m\right)R^{({\cal D})}_{\frac{d}{2}-i\nu \Delta_3 \Delta_4}\left(m,m_{3,4}\right)\\ \times  {\cal A}^{\text{AdS}}_{\Delta_1,\Delta_2|\frac{d}{2}+i\nu|\Delta_3,\Delta_4}\left(Q_1,Q_2,Q_3,Q_4\right)\,,
\end{multline}
and
\begin{multline}
  I_{{\cal A}_-\,{\cal D}_-} = e^{-\frac{d\pi i}{2}}i^{\Delta_1+\Delta_2+\Delta_3+\Delta_4}\left(\prod^4_{i=1}c^{\text{dS-AdS}}_{\Delta_i}\right) \int^{\infty}_{-\infty} \frac{d\nu}{2\pi} \,\,e^{-\left(\frac{d}{2}+i\nu\right)\pi i}\, c^{\text{dS-AdS}}_{\frac{d}{2}+i\nu}\, \\ \times R^{({\cal A})}_{\Delta_1 \Delta_2  \frac{d}{2}+i\nu}\left(m_{1,2},m\right)R^{({\cal D})}_{\frac{d}{2}-i\nu \Delta_3 \Delta_4}\left(m,m_{3,4}\right)\\ \times {\cal A}^{\text{AdS}}_{\Delta_1,\Delta_2|\frac{d}{2}+i\nu|\Delta_3,\Delta_4}\left(Q_1,Q_2,Q_3,Q_4\right).
\end{multline}
Like for the contact diagram in the previous section, using that \eqref{RAtoRD}
\begin{equation}
    R^{({\cal A})}_{\Delta_1 \Delta_2  \frac{d}{2}+i\nu}\left(m_1,m_2,m\right) = e^{-\left(d+2\right)\frac{\pi i}{2}}R^{({\cal D})}_{\Delta_1 \Delta_2  \frac{d}{2}+i\nu}\left(m_1,m_2,m\right),
\end{equation}
one finds the two contributions to be equal and opposite:
\begin{equation}
    I_{{\cal A}_-\,{\cal D}_+}+I_{{\cal D}_-\,{\cal D}_+}=0.
\end{equation}
After having observed this property for contact diagrams and now also for particle exchanges, one understands that the pairwise cancellation of contributions from regions ${\cal A}_-$ and ${\cal D}_-$ is a general property of celestial correlation functions \eqref{ccdefn}, at least in perturbation theory. The proof follows simply from the cancellation \eqref{AmDm0cont} of the contributions from regions ${\cal A}_-$ and ${\cal D}_-$ at the level of contact diagrams, combined with the property \eqref{DmAmsigma} of bulk-to-bulk propagators with points in these regions.

\vskip 4pt
The full celestial exchange diagram \eqref{celestialexch} is therefore the sum of contributions \eqref{Iaa} - \eqref{Idd} from regions ${\cal A}_+$ and ${\cal D}_+$, giving\footnote{To write the exchange in this form in terms of flat-AdS coefficients \eqref{flatads2pt} and \eqref{adstoflatnpt} we used that
\begin{equation}
    R^{({\cal D})}_{\frac{d}{2}-i\nu \, \Delta_3 \Delta_4}\left(m,m_{3,4}\right)=\left(\frac{m}{2}\right)^{2i\nu}\frac{\Gamma\left(-i\nu\right)}{\Gamma\left(i\nu\right)}R^{({\cal D})}_{\frac{d}{2}+i\nu \, \Delta_3 \Delta_4}\left(m,m_{3,4}\right).
\end{equation}} 
\begin{multline}\label{celesexchresult}
   \hspace*{-1cm} \left(-ig_{12}\right)\left(-ig_{34}\right) \int d^{d+2}X\,\int d^{d+2}Y\,G^{(m_1)}_{\Delta_1}\left(X,Q_1\right)G^{(m_2)}_{\Delta_2}\left(X,Q_2\right)G^{\left(m\right)}_{T}\left(X,Y\right)G^{(m_3)}_{\Delta_3}\left(Y,Q_3\right)G^{(m_4)}_{\Delta_4}\left(Y,Q_4\right)\\
  \hspace*{-0.1cm}  = g_{12}\,g_{34} \int^{\infty}_{-\infty} \frac{d\nu}{2\pi} \,\frac{c^{\text{flat-AdS}}_{\Delta_1 \Delta_2 \frac{d}{2}+i\nu}\left(m_1,m_2,m\right)c^{\text{flat-AdS}}_{\frac{d}{2}+i\nu \Delta_3 \Delta_4}\left(m,m_3,m_4\right)}{c^{\text{flat-AdS}}_{\frac{d}{2}+i\nu}} \\ \times {\cal A}^{\text{AdS}}_{\Delta_1,\Delta_2|\frac{d}{2}+i\nu|\Delta_3,\Delta_4}\left(Q_1,Q_2,Q_3,Q_4\right).
\end{multline}
This result expresses the celestial exchange diagram as a continuum of exchange Witten diagrams in EAdS$_{d+1}$ where the exchanged particles have scaling dimension belonging to the Principal Series of the Euclidean conformal group. The coefficient of each exchange Witten diagram is factorised, where the two factors $c^{\text{flat-AdS}}_{\Delta_1 \Delta_2 \frac{d}{2}+i\nu}\left(m_1,m_2,m\right)$ and $c^{\text{flat-AdS}}_{\frac{d}{2}+i\nu \Delta_3 \Delta_4}\left(m,m_3,m_4\right)$ serve to convert the coefficients of the three-point contact subdiagrams in EAdS to their Celestial counterparts via \eqref{adstoflatnpt}, while $c^{\text{flat-AdS}}_{\frac{d}{2}+i\nu}$ accounts for the change in two-point function coefficient \eqref{flatads2pt}.

\vskip 4pt
The expression \eqref{celesexchresult} for the exchange is the celestial analogue of the formula \cite{Sleight:2020obc,Sleight:2021plv} for exchange contributions to late-time correlation functions in dS$_{d+1}$ space, where they are similarly expressed in terms of corresponding exchange Witten diagrams in EAdS$_{d+1}$ with coefficients ensuring consistent factorisation. A key difference is that the result for the dS exchange is a sum of two exchange Witten diagrams for a particle of the same mass in EAdS (one for each boundary condition), while celestial exchanges are a continuum of exchanged particles in EAdS carrying Principal Series representations. This difference is simply owing to the fact that the equations of motion for a field in dS$_{d+1}$ are identified with those of a field of the same mass in EAdS$_{d+1}$ under the analytic continuations \eqref{todsads}. In Minkowski space instead, the radial reduction of a massive scalar in $\mathbb{M}^{d+2}$ onto a single $\left(d+1\right)$-dimensional leaf of the hyperbolic foliation gives rise to a continuum \eqref{spectralrepcelesybubu} of scalar fields carrying Principal Series representations of $SO\left(d+1,1\right)$. From a group theoretical perspective, the appearance of the whole principal series is not a surprise due to the Plancherel theorem for the Lorentz group. It is interesting to note the absence of discrete Complementary Series representations which, as will be discussed in section \ref{subsec::RR}, seems to carry over at the non-perturbative level. 

\vskip 4pt
While in this work we focus on recasting perturbative celestial correlators terms of EAdS Witten diagrams, like the latter they also admit an expansion in Conformal Partial Waves. This follows from harmonic analysis of the Euclidean Conformal Group and will be presented in an upcoming work \cite{toappear}. 

\subsection{Rules to write Celestial Correlators as EAdS Witten diagrams}
\label{subsec::pertcalcrules}

Having seen some explicit examples in the previous sections, we can write down a set of rules to express any given perturbative contribution to celestial correlation functions as defined by \eqref{ccdefn} in terms of Witten diagrams in EAdS:\footnote{In the following rules, for simplicity we will focus on non-derivative interaction vertices and scalar field theories.}

\begin{itemize}
    \item \emph{External lines:} For each celestial bulk-to-boundary propagator $G^{\left(m_i\right)}_{\Delta_i}\left(X,Q_i\right)$, write down the corresponding bulk-to-boundary propagator in EAdS: $K^{\text{AdS}}_{\Delta_i}\left(s_{\text{AdS}}({\hat X},Q_i)-i\epsilon\right)$.
    \item \emph{Internal lines:} For each Minkowski Feynman propagator, write down the bulk-to-bulk propagator in EAdS for a particle with (dummy variable) scaling dimension $\Delta$ and divide by $c^{\text{flat-AdS}}_{\Delta}$. The EAdS bulk-to-bulk propagator associated to each internal line is labelled by a distinct scaling dimension.
    \item \emph{Internal points:} For each internal point, integrate over EAdS and multiply by the corresponding coefficient \eqref{adstoflatnpt} which converts the contact sub-diagram for the EAdS process to its celestial counterpart via \eqref{adstoflatcontact}.
    \item Integrate over the scaling dimensions of each EAdS bulk-to-bulk propagator.   
\end{itemize}

\vskip 4pt
For example, using the above rules we can immediately decompose the four-point ``Candy diagram" as a continuum of such diagrams in EAdS. Consider the diagram generated by quartic vertices: 
\begin{equation}
    {\cal V}_{12 \phi \chi} = g_{12}\, \phi_1 \phi_2 \phi\, \chi, \qquad {\cal V}_{34 \phi \chi} = g_{34}\, \phi_3 \phi_4 \phi \,\chi,
\end{equation}
where $\phi_i$ are external and $\phi,\,\chi$ exchanged with masses $m_{\phi}$ and $m_{\chi}$ and couplings $g_{12}$ and $g_{34}$. 

\vskip 4pt
Using the above rules, the corresponding contribution to celestial correlators is given by 
\begin{multline}
   \hspace*{-0.5cm} \left(-ig_{12}\right)\left(-ig_{34}\right) \int d^{d+2}X\,\int d^{d+2}Y\,G^{(m_1)}_{\Delta_1}\left(X,Q_1\right)G^{(m_2)}_{\Delta_2}\left(X,Q_2\right)G^{\left(m_\phi\right)}_{T}\left(X,Y\right)G^{\left(m_\chi\right)}_{T}\left(X,Y\right)\\ \times G^{(m_3)}_{\Delta_3}\left(Y,Q_3\right)G^{(m_4)}_{\Delta_4}\left(Y,Q_4\right)\\
   = g_{12}\,g_{34} \int^{\infty}_{-\infty} \frac{d\nu_\phi}{2\pi} \frac{d\nu_\chi}{2\pi} \,\frac{c^{\text{flat-AdS}}_{\Delta_1 \Delta_2 \frac{d}{2}+i\nu_\phi \frac{d}{2}+i\nu_\chi}\left(m_1,m_2,m_\phi,m_\chi\right)c^{\text{flat-AdS}}_{\frac{d}{2}+i\nu_\chi\frac{d}{2}+i\nu_\phi \Delta_3 \Delta_4}\left(m_\chi,m_\phi,m_3,m_4\right)}{c^{\text{flat-AdS}}_{\frac{d}{2}+i\nu_\phi}c^{\text{flat-AdS}}_{\frac{d}{2}+i\nu_\chi}}\\ \times  {\cal A}^{\text{AdS}}_{\Delta_1,\Delta_2|\frac{d}{2}+i\nu_\phi \frac{d}{2}+i\nu_\chi|\Delta_3,\Delta_4}\left(Q_1,Q_2,Q_3,Q_4\right),
\end{multline}
in terms of the corresponding Candy Witten diagram in EAdS:
\begin{multline}
    {\cal A}^{\text{AdS}}_{\Delta_1,\Delta_2|\frac{d}{2}+i\nu_\phi \frac{d}{2}+i\nu_\chi|\Delta_3,\Delta_4}\left(Q_i\right)=\int_{{\cal A}_+} d^{d+1}{\hat X}\,\int_{{\cal A}_+} d^{d+1}{\hat Y}\,G^{\text{AdS}}_{\frac{d}{2}+i\nu_\phi}({\hat X},{\hat Y})G^{\text{AdS}}_{\frac{d}{2}+i\nu_\chi}({\hat X},{\hat Y})\\ \times \prod^2_{i=1}K^{\text{AdS}}_{\Delta_i}\left(s_{\text{AdS}}({\hat X},Q_i)-i\epsilon\right)\prod^4_{i=3}K^{\text{AdS}}_{\Delta_i}\left(s_{\text{AdS}}({\hat Y},Q_i)-i\epsilon\right).
\end{multline}

\vskip 4pt
An analogous set of rules were given in \cite{Iacobacci:2022yjo} (section 4.2), which were to express celestial amplitudes as defined in \cite{Pasterski:2016qvg,Pasterski:2017kqt} in terms of corresponding Witten diagrams in EAdS. A difference with respect to the rules given above is that for celestial correlators \eqref{ccdefn} we do not need to distinguish boundary points on $S^+_d$ and $S^-_d$, since the corresponding propagators are given by the same analytic function. In the prescription \cite{Pasterski:2016qvg,Pasterski:2017kqt}, the conformal primary wavefunctions for incoming and outgoing particles have different $i\epsilon$ prescriptions.

\section{Radial reduction of the K\"all\'en-Lehmann spectral representation}
\label{sec::RRKL}

In the previous sections we have seen how to connect QFT in Minkowski space with QFT on the extended hyperboloid via radial reduction. In the following we will use this connection to study the radial reduction of the K\"all\'en-Lehmann representation of the two-point function in Minkowski space. This analysis will allow us to begin the task of importing techniques and methods from non-perturbative QFT to Celestial Correlators. In this work we will limit ourselves to presenting the set-up and some simple examples, where further details and results will be given in an upcoming work \cite{toappear}.

\subsection{Review: K\"all\'en-Lehmann spectral representation}
\label{subsec::revKL}

An important result in non-perturbative QFT is the K\"all\'en-Lehmann spectral representation for exact two-point functions, which in the time-ordered case reads:
\begin{align}
\Gamma(X,Y)&=\langle \Omega | T\left\{\phi(X)\phi(Y)\right\}|\Omega \rangle,\\
&= \int^\infty_0 d\mu^2\, \rho(\mu^2) \,G^{(\mu)}_T(X,Y), \label{KLMink}
\end{align}
or equivalently in momentum space 
\begin{align}
    \Gamma(P^2) &= \int d^{d+2}X \, \Gamma(X,Y)\, e^{i P \cdot (X-Y)}\,,\\ 
    &=\int^\infty_0 d\mu^2\, \rho(\mu^2) \,G^{(\mu)}_T(P)\,,
\end{align}
with positive definite spectral density,
\begin{equation}
    \rho(\mu^2) \geq 0.
\end{equation}
The K\"all\'en-Lehmann representation is is a simple consequence of completeness and unitarity.\footnote{For the time-ordered two-point function one also needs reality of the spectral function $\rho(\mu^2)$, which follows from CPT.} The analytic structure of $\Gamma(P^2)$ is encoded by the spectral density, which takes following the generic form 
\begin{align}
    \rho(\mu^2)\sim Z\delta(\mu^2-m^2)+\theta(\mu^2-\mu^2_0)\sigma(\mu^2)\,.
\end{align}
On the principal (physical) sheet there are no other singularities other than on the positive real axis which are dictated by the spectrum. Analytic continuation then follows from the $i\epsilon$ prescription $-P^2+i\epsilon$.

\vskip 4pt
From the knowledge of the analytic structure one can derive a dispersion relation for the momentum space two-point function:\footnote{UV divergences can be dealt with by considering subtracted dispersion relations, giving an integral representation for $\Gamma(p^2)-\Gamma(p_0^2)$ where the divergence is removed being common to both terms. Another option is to add counterterms as usual. Once the divergence is removed the spectral function can be extracted.}
\begin{align}
    \Gamma(-P^2+i\epsilon)=\frac1{2\pi i}\oint_\gamma dz\, \frac{\Gamma(z)}{z+P^2-i\epsilon}=\frac1{2\pi i}\int_{z_1}^\infty dz\,\frac{\text{disc}[\Gamma(s)]}{z+P^2-i\epsilon},
\end{align}
with discontinuity
\begin{align}
    \text{disc}[\Gamma(s)]\equiv\Gamma(s+i\epsilon)-\Gamma(s-i\epsilon).
\end{align}
This leads to an efficient way to characterise and extract the spectral function from the two-point function discontinuity:
\begin{align}\label{rhodisc}
    \rho(z)=\frac1{2\pi i}\,\text{disc}[\Gamma(z)]=\frac1{\pi}\,
    \Im [\Gamma(z+i\epsilon)]\,.
\end{align}

\vskip 4pt
In the following we shall explore how these properties of the K\"all\'en-Lehmann spectral representation get translated upon radial reduction to the extended unit hyperboloid.

\subsection{Radial reduction}
\label{subsec::RR}

\vskip 4pt The radial reduction of the K\"all\'en-Lehmann spectral representation \eqref{KLMink} onto the extended unit hyperboloid follows from the radial reduction \eqref{spectralrepcelesybubu} of the Feynman propagator:
\begin{equation}\label{KLRR}
\Gamma(X,Y) = \int_{-\infty}^{+\infty}\frac{d\nu}{2\pi}\,\rho_\nu\left(\sqrt{X^2+i\epsilon},\sqrt{Y^2+i\epsilon}\right)G_{\frac{d}{2}+i\nu}(\sigma_\epsilon)\,,
\end{equation}
with spectral function
\begin{align}
\rho_\nu\left(\sqrt{X^2+i\epsilon},\sqrt{Y^2+i\epsilon}\right)=\int_0^\infty d\mu^2 \rho(\mu^2)\,\mathcal{K}^{(\mu)}_\nu(\sqrt{X^2+i\epsilon})\mathcal{K}^{(\mu)}_{-\nu}(\sqrt{Y^2+i\epsilon})\,,
\end{align}
which takes the form of a multiplicative (Mellin) convolution of the spectral density $\rho(\mu^2)$ with the spectral density \eqref{SDFeyn} for the Feynman propagator.

\vskip 4pt
Since the above radially reduced spectral function depends on the radial direction it is useful to consider its form in Mellin space (following section \ref{subsec::celestFeyn}):
\begin{align}
    \Gamma_{\Delta_1,\Delta_2}({\hat X},{\hat Y})&=\int^\infty_0\, \frac{dt_1}{t_1}\frac{dt_2}{t_2}t^{\Delta_1}_1t^{\Delta_2}_2\, \Gamma(t_1 X,t_2 Y),\\
    &=\mathcal{M}[\rho]\left(\tfrac{d-\Delta_1-\Delta_2}2+1\right)\,G^{(m=1)}_{\Delta_1 \Delta_2}({\hat X}_1,{\hat X}_2),
\end{align}
which, as per the Mellin convolution theorem, is the product of the Mellin transform of the spectral density $\rho(\mu^2)$ and of the Feynman propagator \eqref{bubucelestM}:
\begin{align}\label{MTrho}
\mathcal{M}[\rho](s)=\int_0^\infty \frac{d z}{z}\,\rho(z)\,z^{s}\,.
\end{align}
The K\"all\'en-Lehmann spectral representation for the Mellin transformed two-point function then simply follows from that of the Feynman propagator \eqref{spectralMT2pt}:
\begin{align}\label{KLMT}
    \Gamma_{\Delta_1,\Delta_2}({\hat X},{\hat Y})=\int_{-\infty}^{+\infty}\frac{d\nu}{2\pi}\,\rho_{\Delta_1,\Delta_2}\left(\nu\,\right)G_{\frac{d}{2}+i\nu}(\sigma_\epsilon),
\end{align}
with spectral function  
\begin{align}\label{hypspect}
  \rho_{\Delta_1+\frac{d}{2},\Delta_2+\frac{d}{2}}\left(\nu\right)&=\mathcal{M}[\rho](-\tfrac{\Delta_1+\Delta_2}2+1)\,\rho^{(m=1)}_{\Delta_1+\frac{d}{2},\Delta_2+\frac{d}{2}}(\nu),\\
  &=\mathcal{M}[\rho](-\tfrac{\Delta_1+\Delta_2}2+1)\frac{2^{\Delta_1+\Delta_2}}{16\pi\left(\sqrt{\hat{X}^2+i\epsilon}\right)^{\Delta_1+\frac{d}{2}}\left(\sqrt{\hat{Y}^2+i\epsilon}\right)^{\Delta_2+\frac{d}{2}}}\\&\hspace{100pt}\times\frac{\Gamma \left(\frac{\Delta_1-i \nu}{2}\right) \Gamma \left(\frac{\Delta_1+i \nu}{2}\right) \Gamma \left(\frac{\Delta_2-i \nu}{2}\right) \Gamma \left(\frac{\Delta_2+i \nu}{2}\right)}{\Gamma(i\nu)\Gamma(-i\nu)}\,.\nonumber
\end{align}

\vskip 4pt
The Mellin transform is a useful tool to study the properties of the radial reduction of the K\"all\'en-Lehmann spectral representation, which we explore in the following. In particular, we discuss the analytic properties of the Mellin transform $\mathcal{M}[\rho](s)$ of the spectral function and show that the radially reduced K\"all\'en-Lehmann spectral representation can be expressed in the form of a of a Sommerfeld-Watson transformation of (suitably analytically continued) $SO(d+1)$ spherical harmonics, which allows to connect with Euclidean space. In section \ref{subsec::KLEX} we will consider some examples in perturbation theory.

\paragraph{Analyticity of $\mathcal{M}[\rho](s)$.} To study the analytic structure of the Mellin transform $\mathcal{M}[\rho](s)$ of the spectral function $\rho(\mu^ 2)$, we shall assume that the spectral function admits an asymptotic expansion around $0$ and $\infty$ of the form:\footnote{This asymptotic expansion can be generalised to include integer powers of $\log z$ without loss of generality, so for simplicity will not consider such logarithmic terms in the following discussion.} 
\begin{align}
    \rho(z)\Big|_{z\to0}&=\sum_{n}a_n \, z^{\alpha_n},\\
    \rho(z)\Big|_{z\to\infty}&=\sum_{n}a_n \, z^{\beta_n},
\end{align}
in terms of complex numbers $\alpha_n$ and $\beta_n$ whose real part goes to $\pm \infty$ respectively.\footnote{Note that in the presence of a mass-gap the spectral representation is identically vanishing around $z\to0$ and so the asymptotic expansion around $z\sim0$ will be identically $0$.} Under this assumption one can analytically continue the Mellin transform to the whole complex plane with poles at $s\sim -\alpha_n$ and $s\sim -\beta_n$. Let's assume for simplicity that $\rho(z)\sim\sum\limits_n a_n z^n$. This means that we can regularise $\rho(z)$ by subtracting its first $N$ terms of its asymptotic expansion. In practice we can write (see e.g. chapter 3.2 of \cite{MellinBook}) 
\begin{align}
\mathcal{M}[\rho](s)=\int_0^1 dz\left[\rho(z)-\sum_{n=0}^{N-1}a_n z^n\right]z^{s-1}+\sum_{n=0}^\infty\frac{a_n}{s+n}+\int_1^\infty\rho(z)z^{s-1}dz\,.
\end{align}
The integral in the square bracket is now well behaved for $\Re[s]>-N$, giving us a meromorphic analytic continuation of the initial function. A similar argument allows to regularise the integral around $z\sim\infty$ by performing an asymptotic expansion in $u=\frac1{z}$, so that the Mellin transform of a function whose asymptotic expansion involves a sum of powers of $u$ defines a meromorphic function in the complex plane. A similar argument allows to deal with asymptotic expansions involving positive integer powers of $\log z$ (see \cite{MellinBook}), which give higher order poles of order $n$.

\vskip 4pt In summary, under the assumptions stated above the Mellin transform $\mathcal{M}[\rho](s)$ of spectral functions $\rho(z)$ are meromorphic in $s$. In the presence of a mass gap, the only poles that can arise in perturbation theory are in the limit $z\to\infty$, in which case $\mathcal{M}[\rho](s)$ has poles only to the right of the integration contour. In the absence of a mass-gap instead, the spectral function can develop poles also to the left of the contour. These properties of $\mathcal{M}[\rho](s)$, in turn, imply that the corresponding radial spectral function \eqref{hypspect} is a meromorphic function of the spectral parameter $\nu$.

\paragraph{Continuation to Euclidean space.} The radially reduced K\"all\'en-Lehmann spectral representation \eqref{KLRR} can be written in the form of a Sommerfeld-Watson transformation of $SO(d+1)$ spherical harmonics (analytically continued from the Euclidean sphere $S^{d+1}$) upon using the symmetry properties of the spectral integral in $\nu$. This allows to connect with the formulation of QFT in Euclidean space and, moreover, opens up the possibility to formulate (A)dS theories on the sphere through their ambient space description \cite{Dirac:1936fq,Fronsdal:1978vb,Biswas:2002nk,Taronna:2012gb}. 

\vskip 4pt 
To this end, note that the spectral function \eqref{hypspect} is a symmetric function of $\nu$ and for this reason has poles both in the upper- and lower-half $\nu$-plane. However, owing to the symmetry of $G^{\text{dS}}_{\frac{d}{2}+i\nu}(\sigma_\epsilon)$ in $\nu$, we can replace it with an asymmetric function of $\nu$ with improved analyticity properties in the complex $\nu$-plane i.e. poles either all in the upper-half plane or all in the lower-half plane. This boils down to the application of the following simple trigonometric identity:
\begin{multline}\label{TrigID}
\frac{i\nu}{\pi}\,\Gamma(i\nu)\Gamma(-i\nu)\csc \left(\tfrac{\pi}{2} (\Delta_1+\Delta_2)\right) \sin \left(\tfrac{\pi}{2} (\Delta_1-i \nu )\right) \sin \left(\tfrac{\pi}{2}  (\Delta_2-i \nu )\right)\\+(\nu \to -\nu)=1\,,
\end{multline}
which relates two different sets of zeros to the identity. Inserting this into \eqref{KLMT} and changing the integration variable $\nu \to -\nu$ for one of the two terms gives rise to the following spectral function:
\begin{align}\label{SpectUpper}
    \rho_{\Delta_1+\frac{d}{2},\Delta_2+\frac{d}{2}}\left(\nu\right)&=\frac{2^{\Delta_1+\Delta_2}\mathcal{M}[\rho](-\frac{\Delta_1+\Delta_2}2+1)}{8\pi\left(\sqrt{\hat{X}^2+i\epsilon}\right)^{\Delta_1+\frac{d}{2}}\left(\sqrt{\hat{Y}^2+i\epsilon}\right)^{\Delta_2+\frac{d}{2}}}\\&\hspace{100pt}\times (i\nu)\csc(\tfrac{\pi}2(\Delta_1+\Delta_2))\frac{\Gamma \left(\frac{\Delta_1+i \nu}{2}\right) \Gamma \left(\frac{\Delta_2+i \nu}{2}\right)}{\Gamma \left(1-\frac{\Delta_1-i \nu}{2}\right)\Gamma \left(1-\frac{\Delta_2-i \nu}{2}\right)}\,,\nonumber 
\end{align}
where the poles in $\nu$ now lie above the integration contour only. This in turn gives rise to following the spectral function for the radial reduction \eqref{KLRR} of the K\"all\'en-Lehmann spectral representation
\begin{align}\label{asymmrho}
&\rho_\nu\left(\sqrt{X^2+i\epsilon},\sqrt{Y^2+i\epsilon}\right)=\frac{(-i\nu)}{\pi} \int^\infty_0\,d\mu^2\,\rho(\mu^2)\\ \nonumber & \hspace*{1cm}\times \Big[\theta(t_2-t_1)\left(\sqrt{X^2+i\epsilon}\right)^{-\frac{d}{2}}K_{i\nu}\left(\mu \sqrt{X^2+i\epsilon}\right)\left(\sqrt{Y^2+i\epsilon}\right)^{-\frac{d}{2}}I_{-i\nu}\left(\mu  \sqrt{Y^2+i\epsilon}\right)\\ \nonumber &\hspace*{1.2cm}+\theta(t_1-t_2)\left(\sqrt{Y^2+i\epsilon}\right)^{-\frac{d}{2}}K_{i\nu}\left(\mu \sqrt{Y^2+i\epsilon}\right)\left(\sqrt{X^2+i\epsilon}\right)^{-\frac{d}{2}}I_{-i\nu}\left(\mu \sqrt{X^2+i\epsilon}\right)\Big],
\end{align}
upon evaluating the inverse Mellin transform in $\Delta_{1,2}$.

\vskip 4pt
The connection with the Sommerfeld-Watson transformation comes from the following relation between $G_{\frac{d}{2}+i\nu}(\sigma_\epsilon)$ and the Gegenbauer function:
\begin{align}\label{Gidentity}
    G_{\frac{d}{2}+i\nu}(\sigma_\epsilon)=\frac{\Gamma\left(\frac{d}{2}\right)}{4\pi^{d/2}}\frac{C^{(d/2)}_{-\frac{d}{2}-i\nu}(-\hat{X}\cdot\hat{Y})}{\sin\pi \left(\tfrac{d}{2}+i\nu\right)}\,.
\end{align}
With the asymmetric form \eqref{SpectUpper} for the spectral function in $\nu$, the only poles in the lower-half $\nu$ plane in the radial reduction \eqref{KLRR} of the K\"all\'en-Lehmann spectral representation come from the $\sin\pi (\tfrac{d}{2}+i\nu)$ factor in the denominator of \eqref{Gidentity}. Closing the integration contour in the lower-half plane therefore recovers the expansion in spherical harmonics upon analytic continuation to Euclidean space. The absence of poles for $\nu\in -i\left[0,\tfrac{d}{2}\right]$, which would correspond to Complementary Series representations of the Lorentz group, gives a non-perturbative proof (assuming the integrals are convergent) of the completeness of Principal Series representations for scalar fields - which so far we had observed only perturbatively.

\subsection{Examples}
\label{subsec::KLEX}

\paragraph{Single particle states.} The simplest example to start with is for a single particle state, which has spectral function
\begin{equation}
    \rho(\mu^2)=\delta(\mu^2-m^2).
\end{equation}
This has Mellin transform
\begin{equation}
    \mathcal{M}[\rho](-\tfrac{\Delta_1+\Delta_2}2+1)=m^{-\Delta_1-\Delta_2},
\end{equation}
which trivially reproduces the result \eqref{bubucelestM} for the Mellin transform of the free theory two-point function. 

\vskip 4pt
The symmetric form of the radially reduced spectral function is simply:
\begin{align}\label{sdklrrsymm}
&\rho_\nu\left(\sqrt{X^2+i\epsilon},\sqrt{Y^2+i\epsilon}\right)=\mathcal{K}^{(m)}_\nu(\sqrt{X^2+i\epsilon})\mathcal{K}^{(m)}_{-\nu}(\sqrt{Y^2+i\epsilon}),
\end{align}
though it is instructive to consider its asymmetric form \eqref{asymmrho}:
\begin{align}\label{sdklrrasymm}
&\rho_\nu\left(\sqrt{X^2+i\epsilon},\sqrt{Y^2+i\epsilon}\right)\\ \nonumber
&=\frac{(-i\nu)}{\pi} \Big[\theta(t_2-t_1)\left(\sqrt{X^2+i\epsilon}\right)^{-\frac{d}{2}}K_{i\nu}\left(m \sqrt{X^2+i\epsilon}\right)\left(\sqrt{Y^2+i\epsilon}\right)^{-\frac{d}{2}}I_{-i\nu}\left(m \sqrt{Y^2+i\epsilon}\right)\\ \nonumber &\hspace*{1.2cm}+\theta(t_1-t_2)\left(\sqrt{Y^2+i\epsilon}\right)^{-\frac{d}{2}}K_{i\nu}\left(m \sqrt{Y^2+i\epsilon}\right)\left(\sqrt{X^2+i\epsilon}\right)^{-\frac{d}{2}}I_{-i\nu}\left(m \sqrt{X^2+i\epsilon}\right)\Big].
\end{align}
The latter takes the form of the momentum space bulk-to-bulk propagator \cite{Liu:1998ty} in (EA)dS$_{-d+1}$ for a scalar with shadow scaling dimension $-\frac{d}{2}-i\nu$, where the mass $m$ plays the role of the modulus of boundary momentum and the radial coordinate in Minkowski space the role of the bulk direction in Poincar\'e coordinates of (EA)dS. This shows that, in the decomposition \eqref{celesexchresult} of the celestial exchange diagram into corresponding exchange Witten diagrams in EAdS$_{d+1}$, the contribution from the radial direction is in fact given by the \emph{same} Witten diagram but for the shadow fields in momentum space of EAdS$_{-d+1}$. This extends to exchanges the same observation made in \cite{Sleight:2023ojm} for contact diagrams, which was reviewed in section \ref{subsec::pertcalccontact}.

\vskip 4pt
Interestingly, the above observation implies that going from the symmetric form \eqref{sdklrrsymm} of the spectral function to the asymmetric form \eqref{sdklrrasymm} above simply amounts to the momentum space form of identity \eqref{adsharm} between harmonic functions and bulk-to-bulk propagators in EAdS.

\vskip 4pt
Finally, we note that in this case the radially reduced spectral function is an entire function of the spectral parameter $\nu$, in agreement with the expectation that it is a meromorphic function of $\nu$.

\paragraph{1-loop bubble diagram.} Consider a $\lambda \phi^3$ theory in $d+2$ dimensions. The 1-loop two-point bubble-diagram is given by:
\begin{align}
    i\,\Pi(p^2)=\frac{(i\lambda)^2}{2}\int \frac{d^{d+2}k}{(2\pi)^d}\,\frac{-i}{k^2+m^2-i\epsilon}\frac{-i}{(k+p)^2+m^2-i\epsilon}\,.
\end{align}
In the following we will extract the corresponding spectral function $\rho(\mu^2)$ from the discontinuity \eqref{rhodisc} and determine its Mellin transform \eqref{MTrho}.

\vskip 4pt
The momentum integral can be evaluated in the usual way using Feynman parameterisation (see appendix \ref{app::bubble}), arriving to:
\begin{align}
\Pi(p^2)&=\frac{\lambda^2}2\frac{1}{32\pi^{\frac{d+1}2}}\left(\frac{m}{2}\right)^{d-2}\int^{+i\infty}_{-i\infty}\frac{ds}{2\pi i}\,\frac{\Gamma(s)\Gamma(1-s)\Gamma(1-\tfrac{d}{2}-s)}{\Gamma(\tfrac32-s)}\left(\frac{p^2}{4m^2}\right)^{-s}\,,\\
&=\frac{\lambda^2}{2}\,\frac1{16\pi^{\frac{d+2}{2}}} \left(\frac{m}{2}\right)^{d-2} \Gamma \left(\frac{2-d}{2}\right) \, _2F_1\left(1,1-\frac{d}{2};\frac{3}{2};-\frac{p^2}{4m^2}\right).
\end{align}
The discontinuity follows from the standard identity:
\begin{multline}
    \frac1{2\pi i}\,\text{disc}\,\left[ _2F_1(a,b;c;x)\right]\\=\frac{ (x-1)^{-a-b+c}\Gamma(c)}{\Gamma (a) \Gamma (b)} \, _2F_1(c-a,c-b;-a-b+c+1;1-x)\,\theta(x-1),
\end{multline}
giving the spectral function:
\begin{align}\label{osfbub}
    \rho(z)=\frac{\lambda^2}{2}\frac{1}{2^{2d+1}\pi^{\frac{d+1}2}} \frac{\left(z -4 m^2\right)^{\frac{d-1}{2}}}{\sqrt{z}}\ \theta \left(\frac{z }{4 m^2}-1\right)\,,
\end{align}
which is the well-known branch cut of particle production for $z \in \left(4m^2,\infty\right)$. The Mellin transform is simply
\begin{align}
    \mathcal{M}[\rho](s)=\frac{(2 m)^{d+2 s-2}}{2^{2d+1} \pi ^{\frac{d+1}{2}}}\frac{\Gamma \left(\frac{d+1}{2}\right) \Gamma \left(1-\frac{d}{2}-s\right)}{\Gamma \left(\frac{3}{2}-s\right)}\,,
\end{align}
which confirms how the analytic continuation in $s$ is meromorphic and with simple poles along $\Re[s]>0$. The spectral function on the extended unit hyperboloid is then
\begin{multline}
    \rho_{\Delta_1+\frac{d}{2},\Delta_2+\frac{d}{2}}\left(\nu\right)=\frac{m^{d-2 \Delta_1-2 \Delta_2}}{2^{d+5} \pi ^{\frac{d+3}{2}}\left(\sqrt{\hat{X}^2+i\epsilon}\right)^{\Delta_1+\frac{d}{2}}\left(\sqrt{\hat{Y}^2+i\epsilon}\right)^{\Delta_2+\frac{d}{2}}} \frac{\Gamma \left(\frac{d+1}{2}\right)\Gamma \left(\frac{\Delta_1+\Delta_2-d}{2}\right)}{\Gamma \left(\frac{\Delta_1+\Delta_2+1}{2}\right)}\\\times\frac{\Gamma \left(\frac{\Delta_1-i \nu}{2}\right) \Gamma \left(\frac{\Delta_1+i \nu}{2}\right) \Gamma \left(\frac{\Delta_2-i \nu}{2}\right) \Gamma \left(\frac{\Delta_2+i \nu}{2}\right) }{\Gamma (-i \nu ) \Gamma (i \nu )}\,.
\end{multline}
which is meromorphic in the $\Delta_i$. The branch cut singularity of the original spectral function \eqref{osfbub} is replaced by a line of poles in $\Delta_1+\Delta_2$.\footnote{For $d$ even this is an infinite line of poles to the left of the integration contour. For $d$ this truncates to a finite number.} A detailed discussion of loop diagrams in this framework, together with their re-summation in the $O(N)$ and Gross-Neveu models at large $N$, will be presented in an upcoming work \cite{toappear}.

\section*{Acknowledgements}

We thank S\'ebastien Malherbe, Francesca Pacifico, Michel Pannier and Paolo Pergola for discussions. The research of LI, CS and MT is partially supported by the INFN initiative STEFI, by PRIN2022BP52A, and by European Union - Next Generation EU. The research of CS was partially supported by the STFC grant ST/T000708/1 of the Mathematical and Theoretical Particle Physics group at Durham University.

\newpage

\begin{appendix}

\section{Representations of the Feynman propagator}

In this appendix we give the derivations of the various representations of propagators for celestial correlation functions presented in this work.

\subsection{Mellin decomposition and boundary two-point function}
\label{app::doublemellinbubu}

In this section we derive the expression \eqref{bubucelestM} for the Feynman propagator in Mellin space, which is defined by the double Mellin transform:
\begin{align}
    G_{\Delta_1,\Delta_2}(\hat{X},\hat{Y})=\int_0^\infty\frac{dt_1}{t_1}\frac{dt_2}{t_2}\,t_1^{\Delta_1}t_2^{\Delta_2}\,G_T(t_1\hat{X},t_2\hat{Y})\,.
\end{align}
At the end of this second we also take the boundary limit, to obtain the boundary-to-boundary two-point function \eqref{celestial2pt}.

\vskip 4pt
The Mellin transform integrals can be evaluated by employing a Schwinger parametrisation of the Feynman propagator (see e.g. equation (C.4) of \cite{Sleight:2023ojm}):
\begin{multline}
G_{\Delta_1,\Delta_2}(\hat{X},\hat{Y})=\frac{1}{4\pi^{\frac{d+2}2}}\left(\frac{m}2\right)^{\frac{d}{2}}\int_{-i\infty}^{+i\infty}\frac{ds}{2\pi i}\,\Gamma(s-\tfrac{d}{4})\\\times\int_0^\infty\frac{dt}{t}t^{s+\frac{d}{4}}\int_0^\infty\frac{dt_i}{t_i}\,t_i^{\Delta_i}\left(\tfrac{m}2\right)^{-2s}i^{-s-\frac{d}{4}}e^{it(t_1\hat{X}-t_2\hat{Y})^2}\,.
\end{multline}
By sending $t\to t/t_1 t_2$, we have 
\begin{multline}
G_{\Delta_1,\Delta_2}(\hat{X},\hat{Y})=\frac{1}{4\pi^{\frac{d+2}2}}\left(\frac{m}2\right)^{\frac{d}{2}}\int_{-i\infty}^{+i\infty}\frac{ds}{2\pi i}\,\Gamma(s-\tfrac{d}{4})\int_0^\infty\frac{dt}{t}t^{s+\frac{d}{4}}\\\times\int_0^\infty\frac{dt_i}{t_i}\,t_i^{\Delta_i-s-\frac{d}{4}}\left(\tfrac{m}2\right)^{-2s}i^{-s-\frac{d}{4}}e^{it\left(\frac{t_1}{t_2}\hat{X}^2+\frac{t_2}{t_1}\hat{Y}^2-2\hat{X}\cdot\hat{Y}\right)}\,.
\end{multline}
Following this with the change of variables
\begin{equation}
q_1=t_1 t_2\,, \qquad q_2=\frac{t_1}{t_2}\,,
\end{equation}
the integration measure becomes: 
\begin{align}
\int_0^\infty\frac{dt_i}{t_i}\,t_i^{\Delta_i-s-\frac{d}{4}}&=\frac12\int_0^\infty\frac{dq_i}{q_i}q_1^{-\frac{d}{4}+\frac{\Delta_1}{2}+\frac{\Delta_2}{2}-s}q_2^{\frac{\Delta_1-\Delta_2}2}\\&=\pi i\delta\left(s+\frac{d}{4}-\frac{\Delta_1}{2}-\frac{\Delta_2}{2}\right)\int\frac{dq_2}{q_2}\,q_2^{\frac{\Delta_1-\Delta_2}2}\,,
\end{align}
where we have performed the $q_1$ integral since the integrand only depends on $q_2$. The Dirac-delta then allows to perform the $s$ integral, giving
\begin{multline}
G_{\Delta_1,\Delta_2}(\hat{X},\hat{Y})=\frac{1}{8\pi^{\frac{d+2}2}}\left(\tfrac{m}2\right)^{d-\Delta_1-\Delta_2}\,\Gamma(\tfrac{\Delta_1+\Delta_2-d}{2})\int_0^\infty\frac{dt}{t}t^{\frac{\Delta_1+\Delta_2}2}\\\times\int_0^\infty\frac{dq_2}{q_2}\,q_2^{\frac{\Delta_1-\Delta_2}2}i^{-\frac{\Delta_1+\Delta_2}2}e^{it\left(q_2\hat{X}^2+\frac{1}{q_2}\hat{Y}^2-2\hat{X}\cdot\hat{Y}\right)}\,.
\end{multline}
The leftover integrals can be evaluated in terms of a Mellin representation. Using
\begin{align}
    e^{i\frac{t}{q_2}\,\hat{Y}^2}&=\int_{-i\infty}^{+i\infty}\frac{ds}{2\pi i}\,\Gamma(s)\left(-i\frac{t}{q_2}\,\hat{Y}^2+\epsilon\right)^{-s}\\&=\int_{-i\infty}^{+i\infty}\frac{ds}{2\pi i}\,\Gamma(s) t^{-s}q_2^s(-i)^{-s}(\hat{Y}^2+i\epsilon )^{-s}\,,
\end{align}
one can then first evaluate the $q_2$ integral, which is a Schwinger parameterisation, giving
\begin{multline}
G_{\Delta_1,\Delta_2}(\hat{X},\hat{Y})=\frac{1}{8\pi^{\frac{d+2}2}}\left(\tfrac{m}2\right)^{d-\Delta_1-\Delta_2}\,\Gamma(\tfrac{\Delta_1+\Delta_2-d}{2})\int_{-i\infty}^{+i\infty}\frac{ds}{2\pi i}\,\Gamma(s)(-i)^{-s+\frac{\Delta_1+\Delta_2}2}(\hat{Y}^2+i\epsilon )^{-s}\\\times\Gamma\left(s+\tfrac{\Delta_1-\Delta_2}2\right)\int_0^\infty\frac{dt}{t}t^{\frac{\Delta_1+\Delta_2}2-s}(-it\hat{X}^2+\epsilon)^{-s-\frac{\Delta_1-\Delta_2}2}e^{-2it\hat{X}\cdot\hat{Y}}\,.
\end{multline}
Similarly for the $t$-integral, obtaining
\begin{align}
G_{\Delta_1,\Delta_2}(\hat{X},\hat{Y})&=\frac{1}{8\pi^{\frac{d+2}2}}\left(\tfrac{m}2\right)^{d-\Delta_1-\Delta_2}\,\Gamma(\tfrac{\Delta_1+\Delta_2-d}{2})\\\nonumber\times&\int_{-i\infty}^{+i\infty}\frac{ds}{2\pi i}\,\Gamma(s)\Gamma\left(s+\tfrac{\Delta_1-\Delta_2}2\right)\Gamma\left(-2s+\Delta_2\right)(-i)^{-2s+\Delta_2}(\hat{Y}^2+i\epsilon )^{-s}\\\nonumber\times&(\hat{X}^2+i\epsilon)^{-s-\frac{\Delta_1-\Delta_2}2}(2i\hat{X}\cdot\hat{Y}+\epsilon)^{2s-\Delta_2}\,.
\end{align}
After shifting $s\to s-\frac{\Delta_1-\Delta_2}4$, this is
\begin{align}
G_{\Delta_1,\Delta_2}(\hat{X},\hat{Y})&=\frac{1}{8\pi^{\frac{d+2}2}}\left(\tfrac{m}2\right)^{d-\Delta_1-\Delta_2}\,\Gamma(\tfrac{\Delta_1+\Delta_2-d}{2})\\\nonumber\times&\int_{-i\infty}^{+i\infty}\frac{ds}{2\pi i}\,\Gamma(s-\tfrac{\Delta_1-\Delta_2}4)\Gamma\left(s+\tfrac{\Delta_1-\Delta_2}4\right)\Gamma\left(-2s+\tfrac{\Delta_1+\Delta_2}2\right)\\\nonumber\times&(\hat{Y}^2+i\epsilon )^{-s+\frac{\Delta_1-\Delta_2}4}(\hat{X}^2+i\epsilon)^{-s-\frac{\Delta_1-\Delta_2}4}(-2\hat{X}\cdot\hat{Y}+i\epsilon)^{2s-\tfrac{\Delta_1+\Delta_2}2}\,,
\end{align}
or equivalently
\begin{align}\label{MTBBc}
G_{\Delta_1,\Delta_2}(\hat{X},\hat{Y})&=\frac{1}{8\pi^{\frac{d+2}2}}\frac{\left(\tfrac{m}2\right)^{d-\Delta_1-\Delta_2}}{\left(\sqrt{\hat{X}^2+i\epsilon}\right)^{\Delta_1}\left(\sqrt{\hat{Y}^2+i\epsilon}\right)^{\Delta_2}}\,\Gamma(\tfrac{\Delta_1+\Delta_2-d}{2})\\\nonumber\times&\int_{-i\infty}^{+i\infty}\frac{ds}{2\pi i}\,\Gamma(s-\tfrac{\Delta_1-\Delta_2}4)\Gamma\left(s+\tfrac{\Delta_1-\Delta_2}4\right)\Gamma\left(-2s+\tfrac{\Delta_1+\Delta_2}2\right)\\\nonumber&\hspace{100pt}\times\left(\frac{-2\hat{X}\cdot\hat{Y}+i\epsilon}{\sqrt{\hat{X}^2+i\epsilon}\sqrt{\hat{Y}^2+i\epsilon}}\right)^{2s-\tfrac{\Delta_1+\Delta_2}2}\,.
\end{align}
Finally performing the Mellin integral closing on the poles on the positive real axis we arrive to the closed form expression (\emph{cf.} equation \eqref{MBhyp1}):
\begin{align}\label{Gdeltaapp}
G_{\Delta_1,\Delta_2}(\hat{X},\hat{Y})&=\frac{1}{2}\frac{1}{(4\pi)^{\frac{d+1}2}}\frac{m^{d-\Delta_1-\Delta_2}}{\left(\sqrt{\hat{X}^2+i\epsilon}\right)^{\Delta_1}\left(\sqrt{\hat{Y}^2+i\epsilon}\right)^{\Delta_2}}\\&\times\frac{\Gamma(\tfrac{\Delta_1+\Delta_2-d}{2})
 \Gamma (\Delta_1) \Gamma (\Delta_2)}{\Gamma\left(\frac{\Delta_1+\Delta_2+1}{2}\right)} \, _2F_1\left(\begin{matrix}\Delta_1,\Delta_2\\\frac{\Delta_1+\Delta_2+1}{2}\end{matrix};\frac{1-\frac{-\hat{X}\cdot\hat{Y}+i\epsilon}{\sqrt{\hat{X}^2+i\epsilon}\sqrt{\hat{Y}^2+i\epsilon}}}{2}\right)\,.\nonumber
\end{align}

\vskip 4pt
The free theory boundary two-point function is then the boundary limit of the Mellin transformed Feynman propagator \eqref{Gdeltaapp}:
\begin{align}
     G^{(m)}_{\Delta_1 \Delta_2}(Q_1,Q_2)&:= \lim_{{\hat X}_i \to Q_i}\,G^{(m)}_{\Delta_1 \Delta_2}({\hat X}_1,{\hat X}_2).
\end{align}
To this end, it is useful to employ the Mellin-Barnes representation \eqref{MTBBc} of the Gauss hypergeometric function where, since $Q^2_i=0$, we close the contour on the negative real axis, on the poles: 
\begin{align}
    s&=\frac{\Delta_1-\Delta_2}{4},\,\frac{\Delta_1-\Delta_2}{4}-1,\,\frac{\Delta_1-\Delta_2}{4}-2,\,\ldots\,,\\
    s&=\frac{\Delta_2-\Delta_1}{4},\,\frac{\Delta_2-\Delta_1}{4}-1,\,\frac{\Delta_2-\Delta_1}{4}-2,\,\ldots\,.
\end{align}
This gives
\begin{multline}
     G^{(m)}_{\Delta_1 \Delta_2}(Q_1,Q_2)= \frac{1}{8\pi^{\frac{d+2}{2}}}\left(\frac{m}{2}\right)^{d-\Delta_1-\Delta_2}\Gamma\left(\frac{\Delta_1+\Delta_2-d}{2}\right)\\
     \times \left[\frac{\Gamma\left(\Delta_2\right)}{\left(-2Q_1 \cdot Q_2+i\epsilon\right)^{\Delta_2}}\lim_{{\hat X}\to Q_1}\Gamma\left(\frac{\Delta_1-\Delta_2}{2}\right)\left(\sqrt{{\hat X}^2+i\epsilon}\right)^{\Delta_2-\Delta_1}\right.\\ \left.+\frac{\Gamma\left(\Delta_1\right)}{\left(-2Q_1 \cdot Q_2+i\epsilon\right)^{\Delta_1}}\lim_{{\hat Y}\to Q_2}\Gamma\left(\frac{\Delta_2-\Delta_1}{2}\right)\left(\sqrt{{\hat Y}^2+i\epsilon}\right)^{\Delta_1-\Delta_2}\right]\\
     + \ldots\,,
\end{multline}
where the $\ldots$ are subleading in the boundary limit. Regarded as a distribution in $\Delta_1$ or $\Delta_2$ only one of the two terms contributes, giving
\begin{align}
     G^{(m)}_{\Delta_1 \Delta_2}(Q_1,Q_2)=\frac{C^{\text{flat}}_{\Delta_1}}{(-2Q_1\cdot Q_2+i\epsilon)^{\Delta_1}}(2\pi )\delta(i(\Delta_1-\Delta_2))\,, \nonumber
\end{align}
with 
\begin{equation}
   C^{\text{flat}}_{\Delta}= \left(\frac{m}2\right)^{d-2\Delta}\frac{1}{4\pi^{\frac{d+2}2}}\,\Gamma(\Delta)\Gamma(\Delta-\tfrac{d}2).
\end{equation}
 
 \vskip 4pt
Another way to obtain the same result is to take the boundary limit of the radial Mellin transform of the celestial bulk-to-boundary propagator \eqref{celestialbubo},
\begin{equation}
    G^{(m)}_{\Delta_1 \Delta_2}(Q_1,Q_2)=\lim_{{\hat Y}\to Q_2}\int^{\infty}_0\frac{dt}{t}t^{\Delta_2}G^{\left(m\right)}_{\Delta_1}\left(t{\hat Y},Q_1\right),
\end{equation}
as presented in \cite{Sleight:2023ojm}.

\subsection{Spectral representation}
\label{app::spectralrep}

In this section we present a derivation of the spectral representation \eqref{spectralMT2pt} of the Mellin transformed Feynman propagator \eqref{Gdeltaapp}. This makes use of harmonic analysis on EAdS \cite{Cornalba:2007fs,Penedones:2010ue}, which we review below, which can be extended to points in dS via analytic continuation \cite{Sleight:2019mgd,Sleight:2019hfp,Sleight:2020obc,Sleight:2021plv} and, therefore, also to Minkowski space through its hyperbolic foliation.

\vskip 4pt
Consider functions $F({\hat X},{\hat Y})$ with  ${\hat X},{\hat Y}\, \in$ EAdS$_{d+1}$, which depend only on the chordal distance $r$ between ${\hat X}$ and ${\hat Y}$,
\begin{equation}
    \cosh{r} = - {\hat X} \cdot {\hat Y}.
\end{equation}
If such functions are square integrable
\begin{equation}
    \int_{{\cal A}_+}d^{d+1}{\hat X}\,|F({\hat X},{\hat Y})|^2 = \text{Vol}(S^d)\int^\infty_0\, dr \left(\text{sinh}\,r\right)^d\,|F\left(r\right)|^2 \, < \,\infty,
\end{equation}
they admit a spectral decomposition of the following form in terms of Eigenfunctions \eqref{adsharm} of the AdS Laplacian
\begin{equation}
    F\left(r\right) = \int^{\infty}_{-\infty}d\nu\,{\tilde F}\left(\nu\right)\Omega_{\nu}\left(r\right),
\end{equation}
where
\begin{align}
    {\tilde F}\left(\nu\right) &= \frac{1}{\Omega_{\nu}\left(0\right)} \int_{{\cal A}_+}d^{d+1}{\hat X}\,F({\hat X},{\hat Y})\Omega_{\nu}({\hat X},{\hat Y}),\\
    &= \frac{\text{Vol}(S^d)}{\Omega_{\nu}\left(0\right)} \int^\infty_0\, dr \left(\text{sinh}\,r\right)^d\,F(r)\Omega_{\nu}(r),
\end{align}
which follows from completeness and orthogonality of the Eigenfunctions \eqref{adsharm}:
\begin{align}
     \int_{{\cal A}_+}d^{d+1}{\hat Y}\,\Omega_{\nu}({\hat X}_1,{\hat Y})\Omega_{{\bar \nu}}({\hat Y},{\hat X}_2)&=\frac{1}{2}\left[\delta\left(\nu+{\bar \nu}\right)+\delta\left(\nu-{\bar \nu}\right)\right]\Omega_{\nu}({\hat X}_1,{\hat X}_2),\\
     \int^{\infty}_{-\infty}d\nu\,\Omega_{\nu}({\hat X},{\hat Y})&=\delta({\hat X},{\hat Y}).
\end{align}

\vskip 4pt
A well-known example is the spectral representation of the AdS bulk-to-bulk propagator \eqref{bubuads}, which is normalisable for $\Delta > \frac{d}{2}$ and therefore admits a spectral representation of the form:
\begin{equation}
    G^{\text{AdS}}_{\Delta}(r)=\int^{\infty}_{-\infty}d\nu\,g\left(\nu\right)\Omega_{\nu}(r).
\end{equation}
The spectral function is determined by evaluating the inversion formula:
\begin{align}
    g\left(\nu\right)&= \frac{\text{Vol}(S^d)}{\Omega_{\nu}\left(0\right)} \int^\infty_0\, dr \left(\text{sinh}\,r\right)^d\,G^{\text{AdS}}_{\Delta}(r)\,\Omega_{\nu}(r).
\end{align}
In this example, to proceed it is useful to make the change of variables $y=\text{cosh}^{-2}\left(\frac{r}{2}\right)$, so that
\begin{equation}
    g\left(\nu\right)=\frac{\text{Vol}(S^d)}{\Omega_{\nu}\left(0\right)} 2^{d}\int^1_0\,dy\,y^{-\left(\frac{d+3}{2}\right)}\left(\frac{1}{y}-1\right)^{\frac{d-1}{2}}G^{\text{AdS}}_{\Delta}(y)\,\Omega_{\nu}(y).
\end{equation}
By combining this with the following Mellin-Barnes representation of the Gauss hypergeometric function:\begin{equation}\label{MBhyp1}
    {}_2F_1\left(a,b;c;z\right)=\frac{\Gamma\left(c\right)}{\Gamma\left(a\right)\Gamma\left(b\right)} \int^{+i\infty}_{-i\infty}\frac{ds}{2\pi i}\,\frac{\Gamma\left(s+a\right)\Gamma\left(s+b\right)\Gamma\left(-s\right)}{\Gamma\left(s+c\right)}\left(-z\right)^s,
\end{equation}
the integral over $y$ reduces to a ratio of gamma functions (the Beta function)
\begin{equation}
  \int^1_0\,dy\, \left(1-y\right)^{\frac{d+1}{2}-\Delta-s+t-1}\,y^{-d+\Delta+s-t-1} =  \frac{\Gamma \left(\frac{d+1}{2}-\Delta-s+t \right) \Gamma (-d+\Delta+s-t)}{\Gamma \left(\frac{1}{2}-\frac{d}{2}\right)}.
\end{equation}
This gives
\begin{multline}
    g\left(\nu\right)=\frac{1}{\Gamma \left(\frac{1}{2}-\frac{d}{2}\right) \Gamma \left(\frac{d}{2}-i \nu \right) \Gamma \left(\frac{d}{2}+i \nu \right)}\int^{+i\infty}_{-i\infty}\frac{ds}{2\pi i}\,\frac{dt}{2\pi i}\,\frac{\Gamma (-s) \Gamma (-t) \Gamma (s+\Delta )}{\Gamma \left(\frac{d+1}{2}+t\right) \Gamma (-d+s+2 \Delta +1)}\\ \times \Gamma \left(-\tfrac{d}{2}+s+\Delta +\tfrac{1}{2}\right) \Gamma \left(\tfrac{d}{2}+t-i \nu \right) \Gamma \left(\tfrac{d}{2}+t+i \nu \right) \Gamma \left(\tfrac{d}{2}-s+t-\Delta +\tfrac{1}{2}\right) \Gamma (-d+s-t+\Delta ).
\end{multline}
The Mellin-Barnes integrals in $s$ and $t$ both take the form of the second Barnes lemma, which recovers the known result
\begin{align}\nonumber
    g\left(\nu\right)
    &=\frac{1}{\nu^2+\left(\Delta-\frac{d}{2}\right)^2}.
\end{align}

\vskip 4pt
One proceeds in a similar fashion to determine the spectral decomposition of the Mellin transformed Feynman propagator \eqref{bubucelestM}:
\begin{align}
G_{\Delta_1,\Delta_2}(r)=\int_{-\infty}^{+\infty}d\nu\, g_{\Delta_1,\Delta_2}(\nu)\Omega_\nu(r)\,,
\end{align}
where in appendix \ref{app::doublemellinbubu} we showed that
\begin{multline}
    G^{(m)}_{\Delta_1 \Delta_2}(r)=\frac{1}{2}\frac{1}{(4\pi)^{\frac{d+1}2}}\frac{m^{d-\Delta_1-\Delta_2}}{\left(\sqrt{\hat{X}^2+i\epsilon}\right)^{\Delta_1}\left(\sqrt{\hat{Y}^2+i\epsilon}\right)^{\Delta_2}}\\\times\frac{\Gamma(\tfrac{\Delta_1+\Delta_2-d}{2})
 \Gamma (\Delta_1) \Gamma (\Delta_2)}{\Gamma\left(\frac{\Delta_1+\Delta_2+1}{2}\right)} \, _2F_1\left(\begin{matrix}\Delta_1,\Delta_2\\\frac{\Delta_1+\Delta_2+1}{2}\end{matrix};\frac{1-\text{cosh}\left(r\right)}{2}\right).
\end{multline}
The inversion integral takes the following form:\begin{equation}
    g_{\Delta_1,\Delta_2}(\nu)=\frac{\text{Vol}(S^d)}{\Omega_{\nu}\left(0\right)} \int^\infty_0\,dr\,\left(\text{cosh}\left(r\right)+1\right)^{\frac{d}{2}}\left(\text{cosh}\left(r\right)-1\right)^{\frac{d}{2}} G^{(m)}_{\Delta_1 \Delta_2}(r)\,\Omega_{\nu}(r).
\end{equation}
To proceed in this case it is useful to employ a different Mellin-Barnes representation for the Gauss hypergeometric function
\begin{multline}
    {}_2F_1\left(a,b;c;z\right)=\frac{\Gamma\left(c\right)}{\Gamma\left(a\right)\Gamma\left(b\right)\Gamma\left(c-a\right)\Gamma\left(c-b\right)} \\ \times \int^{+i\infty}_{-i\infty}\frac{ds}{2\pi i}\,\Gamma\left(s\right)\Gamma\left(c-a-b+s\right)\Gamma\left(a-s\right)\Gamma\left(b-s\right)\left(1-z\right)^{-s}.
\end{multline}
The integral over $r$ then gives a ratio of gamma functions
\begin{equation}
    \int^\infty_0\,dr\,\left(\text{cosh}\left(r\right)+1\right)^{\frac{d-2(s+t)}{2}}\left(\text{cosh}\left(r\right)-1\right)^{\frac{d}{2}} = \Gamma \left(\frac{d+1}{2}\right)\frac{2^{d-(s+t)}\Gamma (-d+s+t)}{\Gamma \left(\frac{1-d}{2}+s+t\right)},
\end{equation}
so that 
\begin{multline}
     t^{-\Delta_1}_1 t^{-\Delta_2}_2 g_{\Delta_1,\Delta_2}(\nu)= \frac{m^{d-\Delta_1-\Delta_2}\Gamma \left(\frac{d+1}{2}\right) }{\sqrt{{X}^2+i\epsilon}^{\Delta_1}\sqrt{{Y}^2+i\epsilon}^{\Delta_2}} \frac{\cosh (\pi  \nu ) \cos \left(\pi \left( \frac{\Delta_1-\Delta_2}{2}\right)\right)  \Gamma \left(\frac{\Delta_1+\Delta_2-d}{2} \right)}{2\pi ^2 \Gamma \left(\frac{d}{2}-i \nu \right) \Gamma \left(\frac{d}{2}+i \nu \right)}\\ \int^{+i\infty}_{-i\infty}\frac{ds}{2\pi i}\,\frac{dt}{2\pi i}\, \Gamma (-s) \Gamma (-t)\Gamma (s+\Delta_1) \Gamma \left(t+i \nu +\tfrac{1}{2}\right) \Gamma (-t-2 i \nu )\Gamma \left(\tfrac{d}{2}+t+i \nu \right) \\ \times \frac{\Gamma \left(s+\tfrac{\Delta_1-\Delta_2+1}{2} \right) \Gamma (-s-\Delta_1+\Delta_2) \Gamma \left(-\tfrac{d}{2}+s+t+\Delta_1+i \nu \right)}{ \Gamma \left(s+t+\Delta_1+i \nu +\frac{1}{2}\right)}.
\end{multline}
The Mellin-Barnes integral in $t$ can be evaluated using Barnes second lemma
\begin{multline}
   t^{-\Delta_1}_1 t^{-\Delta_2}_2 g_{\Delta_1,\Delta_2}(\nu)=\cos \left(\tfrac{\pi}{2} \left(\Delta_1-\Delta_2\right)\right) \Gamma \left(\tfrac{\Delta_1+\Delta_2-d}{2} \right) \frac{m^{d-\Delta_1-\Delta_2}}{\sqrt{{X}^2+i\epsilon}^{\Delta_1}\sqrt{{Y}^2+i\epsilon}^{\Delta_2}}\int^{+i\infty}_{-i\infty}\frac{ds}{2\pi i}\,\\\frac{\Gamma (-s) \Gamma \left(s+\frac{\Delta_1-\Delta_2+1}{2}\right) \Gamma (-s-\Delta_1+\Delta_2) \Gamma \left(-\frac{d}{2}+s+\Delta_1-i \nu \right) \Gamma \left(-\frac{d}{2}+s+\Delta_1+i \nu \right)}{2\pi  \Gamma \left(-\frac{d}{2}+s+\Delta_1+\frac{1}{2}\right)}.
\end{multline}
For the remaining integral in $s$, closing the integration contour to the left, on the poles
\begin{align}
    s &= 0,\, 1,\,2,\, \ldots ,\\
    s &= \Delta_1-\Delta_2,\,\Delta_1-\Delta_2+1,\,\Delta_1-\Delta_2+2,\,\ldots\,.
\end{align}
yields
\begin{multline}
  t^{-\Delta_1}_1 t^{-\Delta_2}_2 g_{\Delta_1,\Delta_2}(\nu)= \frac{m^{d-\Delta_1-\Delta_2}}{4\sqrt{{X}^2+i\epsilon}^{\Delta_1}\sqrt{{Y}^2+i\epsilon}^{\Delta_2}} \csc \left(\tfrac{\pi}{2}   (\Delta_1-\Delta_2)\right) \Gamma \left(\tfrac{\Delta_1+\Delta_2-d}{2}\right) \\ \times \left[ \frac{\Gamma \left(\tfrac{\Delta_2-\Delta_1+1}{2}\right) \Gamma \left(-\tfrac{d}{2}+\Delta_2-i \nu \right) \Gamma \left(-\tfrac{d}{2}+\Delta_2+i \nu \right)}{\Gamma\left(\tfrac{1-d}{2}+\Delta_2\right)\Gamma\left(-\Delta_1+\Delta_2+1\right)} \, \right. \\ \times {}_3F_2\left(\tfrac{\Delta_2-\Delta_1+1}{2},-\tfrac{d}{2}+\Delta_2-i \nu ,-\tfrac{d}{2}+\Delta_2+i \nu ;\tfrac{1-d}{2}+\Delta_2,-\Delta_1+\Delta_2+1;1\right)\\\left.-\left(\Delta_1 \to \Delta_2\right)\right].
\end{multline}
The sum of two ${}_3F_2$ hypergeometric functions simplifies to a ratio of gamma functions using a three term relation for ${}_3F_2$ (see e.g. §3.5-§3.8 of \cite{Baileybook} and equation (C.26) of \cite{Sleight:2019hfp}), which leads to
\begin{multline}
   R_1^{-\Delta_1-\frac{d}{2}}R_2^{-\Delta_2-\frac{d}{2}}\,g_{\Delta_1+\frac{d}{2},\Delta_2+\frac{d}{2}}(\nu)=\frac{(\tfrac{m}2)^{-\Delta_1-\Delta_2}}{16\pi\left(\sqrt{{X}^2+i\epsilon}\right)^{\Delta_1+\frac{d}{2}}\left(\sqrt{{Y}^2+i\epsilon}\right)^{\Delta_2+\frac{d}{2}}}\\\times\Gamma \left(\frac{\Delta_1-i \nu}{2}\right) \Gamma \left(\frac{\Delta_1+i \nu}{2}\right) \Gamma \left(\frac{\Delta_2-i \nu}{2}\right) \Gamma \left(\frac{\Delta_2+i \nu}{2}\right)\,,\nonumber
\end{multline}
where for symmetry of the expression we shifted $\Delta_i \to \Delta_i +\frac{d}{2}$.

\section{Bubble diagram}
\label{app::bubble}

In this appendix we review the standard calculation of the bubble diagram in $\lambda\phi^3$ theory. After Wick rotation the integral to evaluate is
\begin{align}
    \Pi(k^2)=\frac{\lambda^2}{2}\int \frac{d^dp}{(2\pi)^d}\,\frac{1}{p^2+m^2}\frac{1}{(k+p)^2+m^2}\,.
\end{align}
Introducing a Schwinger parametrisation $\frac1{A}=\int_0^\infty dt e^{-A t}$ for the propagators and performing the Gaussian momentum integration one arrives to:
\begin{align}
    \Pi(k^2)=\frac{\lambda^2\pi^{d/2}}{2}\,\int_0^\infty dt_1 dt_2(t_1+t_2)^{-d/2}e^{-(t_1+t_2)m^2-\frac{t_1 t_2}{t_1+t_2}\,k^2}\,.
\end{align}
Changing integration variables to
\begin{align}
    t_1+t_2&=p\,,\\
    \frac{t_1}{t_2}=q\,,\\
    \frac{t_1t_2}{t_1+t_2}&=\frac{pq}{(1+q)^2}\,,\\
    dt_1dt_2&=\frac{pdpdq}{(1+q)^2}\,,
\end{align}
and using the Mellin representation of the exponential one can write:
\begin{align}
    \Pi(k^2)=\frac{\lambda^2\pi^{d/2}}{2}\,\int\frac{ds}{2\pi i}\,\Gamma(s)\int_0^\infty dpdq \,p^{1-d/2}(1+q)^{-2}e^{-pm^2}\left(\frac{pq}{(1+q)^2}\,k^2\right)^{-s}\,,
\end{align}
so that the $p$ and $q$ integrals can be simply evaluated via
\begin{align}
    \int_0^\infty \frac{dp}{p}\,p^\alpha e^{-p m^2}&=\Gamma (\alpha ) \left(m^2\right)^{-\alpha }\,,\\
    \int_0^\infty dq\, q^\alpha(1+q)^\beta&=\frac{\Gamma (\alpha +1) \Gamma (-\alpha -\beta -1)}{\Gamma (-\beta )}\,.
\end{align}
The final result is then:
\begin{align}
    \Pi(k^2-i\epsilon)=\frac{\lambda^2\pi^{\frac{d+1}2}}{2}\,m^{d-4}\int\frac{ds}{2\pi i}\,\frac{\Gamma(s)\Gamma(1-s)\Gamma(2-\tfrac{d}{2}-s)}{\Gamma(\tfrac32-s)}\,\left(\frac{k^2-i\epsilon}{4m^2}\right)^{-s}\,,
\end{align}
where we have also reinstated the $i\epsilon$ prescription. From the above it is also manifest the pinching singularity for $d\geq 4$.

\end{appendix}

\bibliographystyle{JHEP}
\bibliography{refs}

\end{document}